\documentclass[twocolumn,prd,nofootinbib,superscriptaddress,longbibliography]{revtex4} %

\newcommand\ForInternalReference[1]{}
\newcommand\SkipForEarlyCirculation[1]{}

\newcommand\SkipPP[1]{}
\usepackage[colorlinks=true,citecolor=blue,urlcolor=blue]{hyperref}
\usepackage{verbatim}
\usepackage{graphicx}
\usepackage{dcolumn}
\usepackage{bm}
\usepackage{color}
\usepackage{xcolor}
\usepackage{xspace}
\usepackage{url}
\usepackage{float}
\usepackage{multirow}
\usepackage{adjustbox}
\usepackage[normalem]{ulem}
\usepackage{bm}
\usepackage{placeins}
\usepackage{afterpage}
\usepackage{amsmath,amssymb}
\usepackage{mathtools}
\usepackage{acronym}
\usepackage{pzccal}
\usepackage{physics}
\usepackage{printlen}
\usepackage{subcaption}
\DeclareMathAlphabet{\mathcal}{OMS}{cmsy}{m}{n}
\newcommand\optional[1]{}
\acrodef{NR}[NR]{Numerical Relativity}
\usepackage{color}
\definecolor{amber}{rgb}{1.0, 0.75, 0.0}
\definecolor{orange}{rgb}{1.0, 0.5, 0.0}
\definecolor{amaranth}{rgb}{0.9, 0.17, 0.31}

\graphicspath{{./Figures/}}

\def\ltsima{$\; \buildrel < \over \sim \;$}
\def\simlt{\lower.5ex\hbox{\ltsima}}
\def\gtsima{$\; \buildrel > \over \sim \;$}
\def\simgt{\lower.5ex\hbox{\gtsima}}

\newcommand{\UT}{\affiliation{Center for Gravitational Physics, The University of Texas at Austin, Austin, Texas 78712, USA}}

\begin{document}

\title{Hybridization of second-order gravitational self-force and numerical relativity waveforms for quasi-circular and non-spinning black hole binaries}

\author{H. L. Iglesias}
\thanks{higlesia@utexas.edu}
\UT
\author{L. Durkan}
\UT
\author{D. M. Shoemaker}
\UT

\begin{abstract}
In the past few decades, the waveform community has made advances in producing waveforms that span the inspiral-merger-ringdown of comparable-mass-ratio black hole binaries using  advances in post-Newtonian and numerical relativity (NR) theory along with state-of-the-art gravitational wave models. Current methods in NR have shown progress towards producing stable simulations reaching mass ratios of 1:100; however, the computational cost becomes prohibitively expensive as the mass ratio and the length of the simulation increases. Meanwhile, the gravitational self-force (GSF)  community has developed waveform models that not only generate extreme mass ratio inspiral  waveforms, but also generate near-equal-mass-ratio waveforms with high fidelity. To assess the limits of both the GSF and NR waveforms and alleviate the computational costs of NR, we present hybridized GSF-NR waveforms for non-spinning binary black hole systems in which GSF provides the inspiral, and NR the merger and ringdown. The hybrid waveforms are generated from a set of 68 non-spinning NR waveforms from the SXS catalogue with mass ratios spanning 1:1 to 1:20 and include the (2,2), (2,1), (3,3), (3,2), (4,4), (4,3), and (5,5) spin-weighted spherical harmonic modes. In this paper, we will highlight a selection of these hybrid waveforms and examine the error in the hybridization procedure. We will investigate the impact of subdominant modes on the accuracy of the hybrid waveforms by performing mismatch comparisons with surrogate models. To address the feasibility of hybridizing GSF inspirals with short, high-mass ratio NR waveforms, thereby alleviating computational costs, we will discuss the relationship between mass ratio and the placement of the matching window, which can be used to predict the necessary and optimal number of NR cycles that contribute to the hybrid waveform.
\end{abstract}

\maketitle
\section{Introduction}
\label{sec:intro}

The detection of 218 
gravitational wave (GW) candidates from compact-object binaries \cite{LIGOScientific:2018mvr,LIGOScientific:2021usb,KAGRA:2021vkt, LIGOScientific:2025hdt, LIGOScientific:2025slb} by Advanced LIGO \cite{LIGOScientific:2014pky}, Advanced Virgo \cite{VIRGO:2014yos} and KAGRA \cite{KAGRA:2018plz} in the first three observing runs, and over 200 so far in the fourth \cite{GraceDB}, have provided us with a wellspring of data from which the source properties of binary black holes (BBHs) \cite{LIGOScientific:2016aoc}, neutron star binaries \cite{LIGOScientific:2017vwq}, and neutron star-black hole (BH) binaries \cite{LIGOScientific:2021qlt} can be inferred. These properties include quantities such as the compact objects' masses and spins, as well as the binary's sky position and luminosity distance. Additionally, GWs provide us with the means to impose stringent tests of general relativity (GR) in regions of strong gravity~\cite{LIGOScientific:2016lio,LIGOScientific:2019fpa,LIGOScientific:2020tif,LIGOScientific:2021sio}, and to further our understanding of the formation channels and population distribution of compact-object binaries \cite{LIGOScientific:2025pvj}.

Of the detections made thus far,  candidate events similar to GW190412 \cite{LIGOScientific:2020stg}, with its predicted 30:8 mass ratio, 
call attention to the need to develop waveform models that are faithful to high-mass-ratio systems. Such disparate mass ratio systems are likely to become more common with the advent of future ground~\cite{Punturo:2010zz,Reitze:2019iox} and space-based detectors~\cite{LISA:2017pwj} whose improved low-frequency sensitivity limits will enable the detection of mergers of intermediate-mass BHs~\cite{LISA:2022yao}. The potential to conduct multiband GW astronomy via the joint observation of a system by space- and ground-based GW detectors further stresses the importance of modeling GWs at all mass ratios \cite{2020NatAs...4..260J, PhysRevLett.116.231102}.

The detection and analysis of GWs resulting from the coalescence of compact objects, including the high-mass ratio events of interest here, require accurate waveforms as predicted by GR. Furthermore, these waveforms need to cover the full parameter space of potential sources with  length in GW cycles to cover the full detector frequency band.
Three approaches are used to solve Einstein's equations and predict the waveform: post-Newtonian (PN) theory, an analytic weak-field approximation method suitable for describing the early inspiral phase in which the velocities are assumed to be slow and point-mass-like
\cite{Blanchet_2024}; numerical relativity (NR), a computational method that directly solves the full, non-linear Einstein field equations 
\cite{Baumgarte_Shapiro_2010,2012CQGra..29l4004P}; and the gravitational self-force (GSF) approach, a formalism in black hole perturbation theory which expands the metric about the background of the larger compact object~\cite{2011LRR....14....7P}.

Each of these methods has limitations. The PN approximation loses accuracy during the late inspiral, when the binary separation is small, and where the approximation for point-mass and slow speeds breaks down. While NR can handle modeling binaries in the strong-field regime at close distances, simulations become prohibitively expensive for both large binary separations and large mass ratios. The GSF approach, while effective at modeling waveforms in the extreme-mass-ratio regime, where the perturbation about a single BH background is valid, is assumed to break down at comparable mass ratios. A gap persists between the various theoretical approaches for BBH systems with $q \lesssim 10^3$,  where $q=m_1/m_2 \geq 1$, and $m_1$ and $m_2$ are the binary's component masses.

Despite these challenges, in recent years groups within the NR and GSF communities have made significant strides in their efforts to shrink this gap from both ends. The RIT NR group \cite{2022PhRvD.105l4010H} have explored the intermediate-mass-ratio regime, reporting the computation of waveforms for a sequence of non-spinning BBHs, with mass ratios up to $q=128$, via Zeno's dichotomy approach~\cite{2020PhRvL.125s1102L}. Ref. \cite{2023CQGra..40iLT01L} also  presented a proof of principle that shows NR and the moving puncture formalism \cite{2006PhRvL..96k1101C} can be used to perform BBH simulations of head-on collisions for mass ratios of up to $q=1000$. Moreover, the SXS \cite{SXS} collaboration released their third public catalog containing NR simulations with mass ratios up to $q=20$ \cite{Scheel:2025jct}.

Advancements in GSF theory, meanwhile, have pushed the approach towards comparable mass ratios for non-spinning, quasi-circular BBHs.  Note that the GSF approximate uses powers of the small mass ratio, $\epsilon=1/q \leq 1$, and solves the Einstein field equations order-by-order, from which the dynamics of the coalescing binary and the metric perturbations are obtained \cite{2011LRR....14....7P}. After estimating the ranges of applicability of different approximation methods for modeling non-spinning, quasi-circular BBHs, Ref. \cite{2020PhRvL.125r1101V} showed that the validity of the GSF approach extends to comparable masses when including the first post-adiabatic (PA), or 1PA, contribution to the orbital phase; this builds on work in Ref. \cite{2014IJMPD..2330022L} which demonstrated black hole perturbation theory may find applications in the radiative inspiral of intermediate-mass-ratio and comparable-mass BH binaries.  The 1PA contribution requires knowledge of the full first-order GSF for quasi-circular orbits and the dissipative part of second-order GSF (2GSF) for quasi-circular inspirals \cite{2021PhRvD.103f4048M}. Building on the second-order self-force calculations of the binding energy of a particle around a Schwarzschild BH \cite{2020PhRvL.124b1101P} and GW energy fluxes \cite{2021PhRvL.127o1102W}, Ref. \cite{2023PhRvL.130x1402W} then presented the first 2GSF calculation for non-spinning, quasi-circular waveforms, which uses a time-domain, post-adiabatic (1PAT1) model. In addition to this calculation, Ref. \cite{2023PhRvL.130x1402W} demonstrated that 1PAT1 waveforms match remarkably well in the inspiral regime with NR waveforms for close-to-comparable mass ratios. Similar findings were presented in Ref. \cite{2024ScPP...17...56K}, in which waveforms up to second post-leading transition-to-plunge order were found to be comparable to NR for small $q$ in the transition-to-plunge regime. These results, coupled with the advancements in NR, point to the possibility of constructing a bridge between the NR and GSF approaches through the ``hybridization" of their respective waveforms.

Hybridization involves smoothly stitching analytical waveforms that cover the early inspiral stage to NR waveforms that capture late times. This hybridization procedure allows the construction of inspiral-merger-ringdown (IMR) waveforms  across the full frequency band of GW detectors. In particular, the hybridization of PN waveforms with NR waveforms has led to the construction of surrogate models based on PN-NR hybrids \cite{2019PhRvD..99f4045V,2022PhRvD.106d4001Y,2023PhRvD.108f4027Y}.

Because the methods used to compute waveforms cover different regions of parameter space, the applicability of hybridized waveforms is restricted to regions of overlap; this can lead to limitations in surrogate models built on hybrid waveforms. For example, in several of these hybrid surrogate models, the amplitudes for all waveform modes were obtained from PN due to its accuracy in the inspiral regime for mass ratios up to $q=8$ \cite{2019PhRvD..99f4045V}.  Nevertheless, this method does not work  at higher mass ratios: Fig. 3 in Ref. \cite{2022PhRvD.106d4001Y} shows significant deviations between the amplitudes of PN and NR at $q=15$ for the $(2,2)$, $(2,1)$, $(3,3)$, $(4,4)$, and $(5,5)$ modes; the 1PAT1 model, however, shows no such deviations for these modes at $q=15$.  Therefore, in the case of non-spinning, quasi-circular BBHs, this issue can be overcome by using 1PAT1 waveforms in the early inspiral.

In this work we propose and construct hybrid waveforms of NR and 2GSF for non-spinning, unequal-mass ratio systems of BBHs. Each hybrid waveform consists of a 1PAT1 inspiral waveform that is smoothly stitched to an NR waveform in the late inspiral phase. Our motivation is twofold: we want (1) to provide unequal mass ratio waveforms that cover the full frequency band of current and future GW detectors; and (2) to discover how many GW cycles are required from NR simulations. To further elucidate the second point, if 2GSF inspiral waveforms can be hybridized with NR waveforms close to merger, then this indicates that simulations can be launched at shorter separations, thus significantly decreasing computational time and allowing current NR codes to provide these cycles.

To share our findings according to the points above, we have organized the paper as follows. In Sec. \ref{sec:waveforms} we describe the NR and 2GSF waveforms used to construct the hybrid waveforms. In Sec. \ref{sec:hybrid} we detail the procedure used to build the hybrid waveforms. In Sec. \ref{sec:results} we assess the accuracy of the hybrid waveforms by analyzing the $L_2$-norm errors that arise from the hybridization, in addition to the mismatches with surrogate models. In Sec. \ref{sec:search_opt_window} we investigate the relationship between mass ratio and the optimal location of the hybridization window. Finally, in Sec. \ref{sec:conclusions} we review our results and discuss future work. Throughout this paper, we use geometrized units with $G=1$ and $c=1$. We define the large mass ratio $q=m_1/m_2 \geq 1$, the small mass ratio $\epsilon=1/q \leq 1$, and the symmetric mass ratio $\nu = m_1 m_2/M^2$, where $M=m_1+m_2$ is the total mass. We note that $\nu$ is related to $q$ through $\nu = q / (1+q)^2$.

\section{Waveforms}
\label{sec:waveforms}
We will need to generate waveforms from NR and GSF to build our hybrids. The two methods have some points in common. In general, we represent the two polarizations of a GW—the plus ($h_{+}$) and cross ($h_{\cross}$) polarizations—as a single complex time-series, $h(t)=h_+(t) - i h_{\cross}(t)$. We note that the polarizations are related to the metric perturbation in GSF theory, which consist of radial functions that are then extrapolated to infinity \cite{Barack:2007tm, Barack:2005nr, Durkan_2022}. The complex GW strain $h$ can be decomposed into a sum of spin-weighted spherical harmonic modes $h_{\ell m}$ via
\begin{equation}
    h(t,\theta,\phi) = \frac{M}{r} \sum_{\ell=2}^{\infty} \sum_{m=-\ell}^\ell h_{\ell m} (t) _{-2} Y_{\ell m }(\theta, \phi),
    \label{eq:strain}
\end{equation}
where $_{-2}Y_{\ell m}$ are the spherical harmonics of spin-weight $-2$, $\theta$ and $\phi$ are the polar and azimuthal angles for the direction of GW propagation in the binary's source frame, and $r$ is the distance of the source from the detector. In Eq. (\ref{eq:strain}), the quadrupole modes ($\ell=|m|=2$) dominate the sum; nevertheless, as $q$ increases, the subdominant modes (also known as non-quadrupole or higher-order modes) begin to play an important role in parameter estimation of GW sources \cite{2017PhRvD..96l4024V, 2014PhRvD..90l4004V, 2014PhRvD..89j2003C, 2020PhRvD.101l4054S, 2021PhRvD.104h4068I}. For this reason, our hybrid waveforms include the $(2,2)$, $(2,1)$, $(3,3)$, $(3,2)$, $(4,4)$, $(4,3)$, and $(5,5)$ spin-weighted spherical harmonic modes.  The $(3,1)$, $(4,2)$, $(4,1)$, $(5,4)$, $(5,3)$, $(5,2)$, and $(5,1)$ modes are excluded as explained in Sec. \ref{subsec:modes}. Also note that the negative \textit{m} modes do not need to be modeled separately as they can be derived from the positive \textit{m} modes through the relation $h_{\ell (-m)} = (-1)^\ell h^*_{\ell m}$, where $h^*_{\ell m}$ is the complex conjugate of $h_{\ell m}$, due to the orbital-plane symmetry of non-precessing BBHs.

%$\mathpzc{h}$

\subsection{NR simulations}
\label{subsec:NR}
The NR waveforms we used are from a set of 68 simulations in the SXS \cite{SXS} public catalog \cite{Scheel:2025jct} produced with the Spectral Einstein Code (SpEC) \cite{SpEC}. A list of these simulations can be found in Appendix \ref{appendix:NR}.  These simulations were selected because they described non-spinning, quasi-circular BBHs, matching the restricted parameter space of the 1PAT1 model: the mass ratios for these simulations fall in the range $1 \leq q \leq 20$; the simulations have dimensionless spin magnitudes less than $2 \times 10^{-4}$; and the largest eccentricity of these simulations is $4.03 \times 10^{-3}$ (SXS:BBH:2598), whereas the median eccentricity is $1.91 \times 10^{-4}$. The maximum dimensionless spin magnitudes for the primary and secondary BHs are $8.41 \times 10^{-5}$ (SXS:BBH:2374) and $1.06 \times 10^{-4}$ (SXS:BBH:1222), respectively. We chose to use SXS simulations for this analysis because the extensive duration of their waveforms allows us to test the validity of hybridization.

For each simulation, the GW strain of Eq. (\ref{eq:strain})  has been extrapolated to future null infinity \cite{2009PhRvD..80l4045B}; we use the strain that has been corrected to account for the supervenient motion of the center of mass \cite{2016PhRvD..93h4031B,boyle_2024_12585016}. Additionally, the highest level of numerical resolution available is used for a given simulation. To avoid the impact of junk radiation on the initial masses, we truncate each NR waveform to its reference time, which is provided in the metadata of the corresponding NR simulation \cite{Scheel:2025jct}. Finally, because our hybridization procedure requires higher accuracy during the merger and ringdown of these NR waveforms, we use the lowest extrapolation order, $N=2$. This follows the rule of thumb in Sec. 2.4.1 of Ref. \cite{Boyle:2019kee}, which suggests using higher-order extrapolation for analyses that require higher accuracy in the inspiral.

\subsection{The 1PAT1 model}
\label{subsec:1PAT1_model}
We compute 2GSF waveforms following the approach in Refs. \cite{Durkan_2022, 2023PhRvL.130x1402W,2021PhRvL.127o1102W}. Here we review the 2GSF time-domain waveform model, 1PAT1, an approximate model of the exact 1PA formalism. For a detailed explanation of the model, see Ref. \cite{2022PhRvD.106h4061A}. The 1PAT1 waveforms of Ref. \cite{2023PhRvL.130x1402W} are calculated by solving the Einstein field equations through second order in $\epsilon$ for non-spinning, quasi-circular BBHs. The 1PAT1 model assumes that the larger compact object is a non-spinning BH described by a background metric $g_{\alpha \beta}$, and that the secondary is any non-spinning compact object described by a metric perturbation $h_{\alpha \beta}$. The full metric of the binary, $\mathbf{g}_{\alpha \beta}$, is expanded about the background metric, $g_{\alpha \beta}$, in powers of  $\epsilon$,
\begin{equation}
    \mathbf{g}_{\alpha \beta} = g_{\alpha \beta} + h_{\alpha \beta} = g_{\alpha \beta}(x^\mu) + \sum_{n=0}^{\infty} \epsilon^n h^n_{\alpha \beta}(x^\mu; z^\mu),
    \label{eq:metric_decomp}
\end{equation}
where $g_{\alpha \beta}$ depends on the background coordinates $x^\mu = (t, r, \theta, \phi)$, and $h^{n}_{\alpha \beta}$ is the $n$th-order metric perturbation which depends on both the background coordinates $x^\mu$ and the position  on the secondary's worldline $z^{\mu}(\tau) = (t_p(\tau), r_p(\tau), 0, \phi_p(\tau))$, where the subscript $p$ denotes quantities representing the position of the secondary \cite{PhysRevD.106.084023}.
In this paper, $g_{\alpha \beta}$ is the Schwarzschild metric.

In GSF theory, 
we can consider two different timescales: the radiation-reaction timescale, the time over which the orbital radius $r_0$ shrinks due to emission and absorption of GWs, and the orbital timescale, the time taken to complete an orbit. The orbital radius will slowly evolve over the radiation-reaction timescale, so any other orbital parameters that depend on $r_0$, such as the frequency and amplitude of the GWs, will also slowly evolve. The orbital phase $\phi_p$, however, will rapidly evolve on the orbital timescale \cite{PhysRevD.106.084023}. The disparity in these timescales allows one to employ the two-timescale expansion \cite{2021PhRvD.103f4048M}, in which the $n^{\text{th}}$-order metric perturbation $h^n_{\alpha \beta}$ is further expanded in terms of slowly evolving amplitudes $h_{\alpha \beta}^{(n,m)}$ and rapidly evolving phase factors $e^{-i m \phi_p}$ at each order $\epsilon^n$. References \cite{2008PhRvD..78f4028H} and \cite{2021PhRvD.103f4048M} have shown that the expansion in Eq. (\ref{eq:metric_decomp}) must be carried though order $\epsilon^2$ to ensure that the accumulated error in the orbital phase is much less than 1 radian over the binary's lifetime, a requirement for high-precision parameter estimation by detectors. Under these assumptions, the perturbation $h_{\alpha \beta}$ must therefore be expanded again, through order $\epsilon^2$,

\begin{equation}
    h_{\alpha \beta} = \sum_{m=-\infty}^\infty \left[\epsilon h_{\alpha \beta}^{(1,m)}(J_A,x^i)+\epsilon^2 h_{\alpha \beta}^{(2,m)}(J_A,x^i) \right]e^{-i m \phi_p},
    \label{eq:h_two_timescale}
\end{equation}
where $x^i=(r,\theta,\phi)$ are the spatial coordinates. 
The time dependence is encoded in $J_A = (m_1, s_1, \Omega)$—the primary BH's mass and spin and the secondary's orbital frequency $\Omega = d\phi_p/dt$—and $\phi_p$. Within this framework, the $n^{\text{th}}$ PA 
approximation includes all terms contributing through order $\epsilon^{n+1}$. In the equations that follow, a numerical superscript $n$ indicates a quantity computed from the $n'$-order amplitudes $h_{\alpha \beta}^{(n',m)}$ up to $n'=n$, while a numerical subscript $n$, as seen in Eq. (\ref{eq:orb_freq_evol_eq_2}), 
indicates the PA order at which the quantity contributes. We note that the first-order metric perturbation contributes at 0PA order.

Although the evolution of the primary's mass and spin appears at 1PA order, its effects are numerically subdominant~\cite{2022PhRvD.106h4061A,2023PhRvL.130x1402W, Albertini:2022dmc} and ignored. Therefore, we only need to consider the evolution of the orbital frequency $\Omega$ and $\phi_p$ \cite{2021PhRvL.127o1102W,2023PhRvL.130x1402W}. There are several methods to calculate the evolution of the orbital frequency, 
one of which makes use of the binding energy, the Bondi-Sachs mass-loss formula and a flux-balance law to calculate the phase to 1PA order \cite{2021PhRvL.127o1102W}. At the time of writing, the data for the binding energy and GW fluxes have been calculated, whereas the second-order self-force has not. For this reason, we will use the former in the construction of 1PA waveforms. The evolution of the orbital frequency and phase are calculated by solving the following coupled differential equations \cite{2023PhRvL.130x1402W}
\begin{equation}
    \dv{\Omega}{t} = - \left(\pdv{E_{\text{bind}}}{\Omega}\right)^{-1} \mathcal{F}, \qquad \dv{\phi_p}{t} = \Omega,
    \label{eq:orb_freq_evol_eq_1}
\end{equation}
where $\mathcal{F}$ is the GW energy flux and $E_{\text{bind}}$ is the binding energy defined in Eqs. (\ref{eq:flux}) and (\ref{eq:E_bind}), respectively. Although the flux and binding energy  can be expressed  as expansions in powers of $\epsilon$ at fixed $m_1$, they have been rewritten in terms of $x=(M \Omega)^{2/3}$, a dimensionless measure of the inverse separation of the BHs; the secondary BH's orbital radius $r_0$, whose relationship with $\Omega$ is given in Eq. (\ref{eq:Omega_of_t}); and the symmetric mass ratio $\nu$ at fixed $M$, the latter of which is used to obtain comparisons commensurable with NR. The flux and binding energy are then re-expanded in powers of $\nu$ to restore the system's discrete symmetry under the interchange $m_1 \leftrightarrow m_2$ of the two bodies, leading to better comparisons with NR \cite{2014IJMPD..2330022L}. Using the two-timescale expansion, the flux \cite{2021PhRvL.127o1102W} can be written as
\begin{equation}
    \mathcal{F} = \nu^2 \mathcal{F}_\nu^{1}(x) + \nu^3 \mathcal{F}_\nu^{2}(x) + \mathcal{O}(\nu^4),
    \label{eq:flux}
\end{equation}
where $\mathcal{F}_\nu^{1} = \mathcal{F}_\nu^{1,\infty}+\mathcal{F}_\nu^{1,\mathcal{H}}$ and $\mathcal{F}_\nu^{2} = \mathcal{F}_\nu^{2,\infty}$; the fluxes $\mathcal{F}^{\infty}$ and $\mathcal{F}^{\mathcal{H}}$ are the energy fluxes radiated to infinity and through the horizon, respectively. We note that the second-order flux excludes the second-order horizon flux as it has not been calculated, but this quantity is expected to be numerically small compared to the flux to infinity \cite{2022PhRvD.106h4061A, 2023PhRvL.130x1402W, 2021PhRvL.127o1102W}. The binding energy is similarly expanded, as
\begin{equation}
    E_{\text{bind}} = \nu M [\hat{\mathcal{E}}_0(x) + \nu \hat{E}_{\text{SF}}(x) + \mathcal{O}(\nu^2)],
    \label{eq:E_bind}
\end{equation}
where $ \hat{\mathcal{E}}_0 (x) = (1-2x)/(\sqrt{1-3x}) - 1$ is the leading-order specific binding energy and $\hat{E}_{\text{SF}}$ is the first-order binding energy due to the self-force \cite{2023PhRvL.130x1402W}; we note that the contributions $\hat{\mathcal{E}}_0 \coloneq \mathcal{E}_0/M$ and $\hat{E}_{\text{SF}} \coloneq E_{\text{SF}}/M$ to the binding energy are adimensional. One final approximation is made by evaluating $\hat{E}_{\text{SF}}$ using a result obtained from the first law of binary mechanics  \cite{2012PhRvL.108m1103L, 2012PhRvD..85f4039L, 2020PhRvL.124b1101P},
\begin{align}
    \hat{E}_{\text{SF}} (x) \approx E^{1^{st} \text{law}} &= -\frac{1}{3} x z_{\text{SF}}'(x) + \frac{z_{\text{SF}}(x)}{2} + \frac{(7-24x)x}{6(1-3x)^{3/2}} \nonumber \\ & \qquad  + \sqrt{1-3x} - 1,
    \label{eq:E_SF}
\end{align}
where $z_{\text{SF}}$ is the first-order correction to the redshift, also known as the first-order Detweiler redshift \cite{2012PhRvL.108m1103L, 2012PhRvD..85f4039L, 2020PhRvL.124b1101P}. The redshift $z_{\text{SF}}$ can be interpreted as the proper time derivative of the time component of the secondary's position \cite{2008PhRvD..77l4026D}. We note that a similar formula for $E^{1^{st} \text{law}}$ is reported in Eq. (14) of \cite{2020PhRvL.124b1101P}, where the difference stems from their having expressed all quantities as functions of the physical parameter $y=(M_{BH} \Omega)^{3/2}$ and expanded in powers of $q \coloneq m/M_{BH}$, where $M_{BH}$ is the primary's perturbed mass and $m$ is the mass of the secondary object. Although a formula for the direct calculation of $\hat{E}_{\text{SF}} (x)$ exists (see Eq. (13) of Ref. \cite{2020PhRvL.124b1101P}), the motivation for using the first-law binding energy is purely logistical: data for the redshift, and hence the first-law binding energy $E^{1^{st} \text{law}}$, is readily available, is easier to calculate than $\hat{E}_{\text{SF}} (x)$, and agrees very well with that from GSF results \cite{2020PhRvL.124b1101P}.

After expanding $d\Omega/dt$ in powers of $\nu$, we can rewrite Eq. (\ref{eq:orb_freq_evol_eq_1}) as
\begin{equation}
    \dv{\Omega}{t} = \frac{\nu}{M^2} (F_0 + \nu F_1), \qquad \dv{\phi_p}{t} = \Omega(t),
    \label{eq:orb_freq_evol_eq_2}
\end{equation}
where $F_0 (x) = a(x) \mathcal{F}_\nu^{1}(x)$ and $F_1(x) = a(x) \mathcal{F}_\nu^{2}(x) - a(x)^2 \mathcal{F}_\nu^{1} (x) \partial_{\hat{\Omega}} \hat{E}_{\text{SF}}$, with $\hat{\Omega}= M \Omega$ and 
\begin{equation}
    a(x) = -\left(\pdv{\hat{\mathcal{E}}_0}{\hat{\Omega}}\right)^{-1} = \frac{3x^{1/2} (1-3x)^{3/2}}{1-6x}.
    \label{eq:a_of_x}
\end{equation}
The set of differential equations in Eq. (\ref{eq:orb_freq_evol_eq_2}) are then integrated after defining the orbital frequency as that of a circular orbit,
\begin{equation}
    \Omega(t) = \sqrt{\frac{M}{r_0^3 (t)}},
    \label{eq:Omega_of_t}
\end{equation}
and setting the initial conditions to be 
\begin{equation}
    \Omega = \eval{\sqrt{\frac{M}{r_0^3 (t)}}}_{t=0}, \qquad \phi_p=\eval{0}_{t=0}.
    \label{eq:initial_conditions}
\end{equation}

After decomposing the strain into spin-weighted spherical harmonic $(\ell, m)$ modes, the full GW waveform for a given mode is constructed using
\begin{equation}
    h_{\ell m} = A_{\ell m} (t) e^{-i \Phi_{\ell m}(t)},
    \label{eq:full_hlm}
\end{equation}
where $\Phi_{\ell m} (t) = m\,\phi_p (t)$. The GW amplitude $A_{\ell m}$ is expanded in powers of $\nu$ up to 1PA order using the two-timescale approximation and can be written as
\begin{equation}
    A_{\ell m} = \nu A_{\ell m}^1 + \nu^2 A_{\ell m}^2,
    \label{eq:amp}
\end{equation}
where the $n$PA approximation includes all terms contributing through order $\nu^{n+1}$. The amplitudes $A^{n}_{\ell m}$ at $0$PA and $1$PA order are calculated as follows
% \begin{equation}
%     h_{\ell m} = \left\{ \nu h_{\ell m}^{(1)} + \nu^2 \left[h_{\ell m}^{(1)} + h_{\ell m}^{(2)} -\frac{2x}{3} \dv{h_{\ell m}^{(1)}}{x} \right] \right\}e^{-i m \phi_p(t)}
% \end{equation}
\begin{align}
    A_{\ell m}^1 (t) &= h_{\ell m}^{(1)} (t), \label{eq:A_1_amp} \\ 
    A_{\ell m}^2 (t) &= h_{\ell m}^{(1)} (t) + h_{\ell m}^{(2)} (t) -\frac{2x}{3} \dv{h_{\ell m}^{(1)} (t)}{x},
    \label{eq:A_2_amp}
\end{align}
where $h_{\ell m}^{(n)}$ is the asymptotic amplitude from Eq. (\ref{eq:h_two_timescale}) evaluated at $J_A = (M,0,\Omega)$; and $h_{\ell m}^{(n)} (t) = h_{\ell m}^{(n)} (\Omega_{\text{1PA}}(t))$, where $\Omega_{\text{1PA}}$ is the solution to Eq. (\ref{eq:orb_freq_evol_eq_2}) when including terms up to $\nu^2$. For a closer look at how these amplitudes are calculated, we direct the reader to Appendix \ref{appendix:1PAT1}. Once the orbital frequency and phase have been calculated from Eq. (\ref{eq:orb_freq_evol_eq_2}), they are inserted into Eqs. (\ref{eq:full_hlm}) and (\ref{eq:amp}), allowing us to generate the 1PAT1 waveforms,
\begin{equation}
    h_{\ell m}^{\text{1PA}}(t) = \left [ \nu A_{\ell m}^{1}(t) + \nu^2 A_{\ell m}^{2}(t) \right ]e^{-i m \phi_p(t)}.
    \label{eq:1PAT1_hlm}
\end{equation}

While the 1PAT1 model produces waveforms that agree  well with NR, we note that this is only true for even-$m$ modes. The same cannot be said for odd-$m$ modes, whose amplitudes do not agree with that of NR; this discrepancy is more apparent at close-to-equal mass ratios where the 1PAT1 amplitudes can be either much larger or smaller than NR waveform amplitudes. This problem stems from the model's inability to capture the inherent discrete antisymmetry of the odd-$m$ modes under the interchange $m_1 \leftrightarrow m_2$ of the two bodies. To restore this antisymmetry, we multiply and divide Eq. (\ref{eq:1PAT1_hlm}) by the reduced mass difference $(m_1 - m_2) / M = \sqrt{1-4 \nu}$, expand the denominator in powers of $\nu$, and truncate the result at order $\nu^2$. After applying this resummation, we can generate odd-$m$ modes with improved accuracy using
\begin{equation}
    h_{\ell m}^{\text{1PA}}(t) = \sqrt{1-4\nu}\left [ \nu A_{\ell m}^{1}(t) + \nu^2 \Bar{A}_{\ell m}^{2}(t) \right ]e^{-i m \phi_p(t)},
    \label{eq:1PAT1_hlm_resummed}
\end{equation}
where $\Bar{A}_{\ell m}^{2}(t) = 2 A_{\ell m}^{1}(t) + A_{\ell m}^{2}(t)$.

\begin{figure*}
    \centering
    \includegraphics[width=1.0\linewidth]{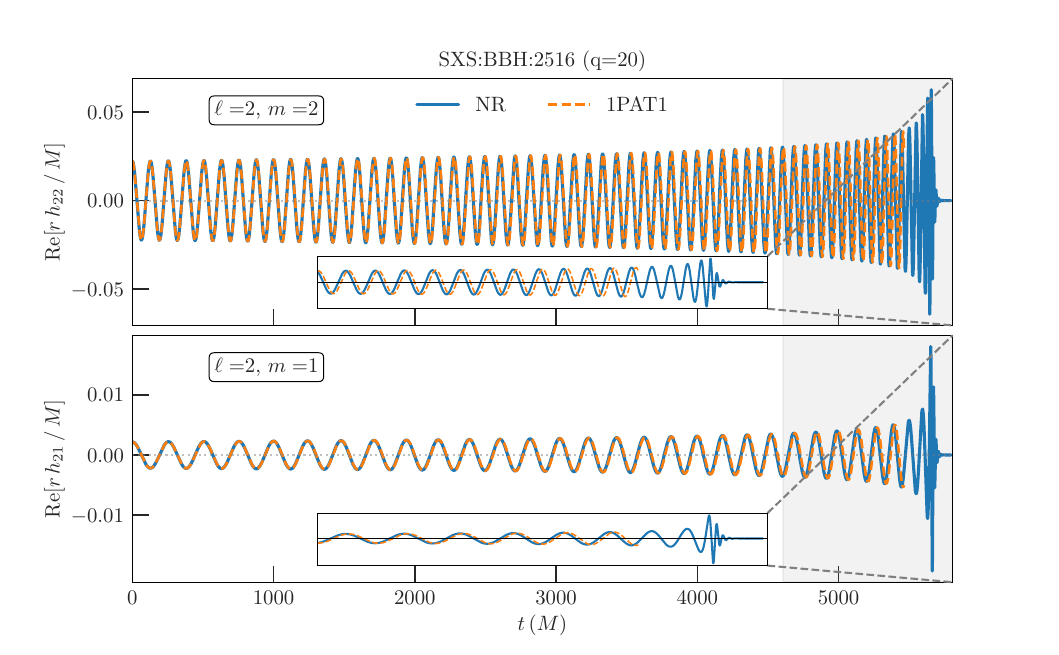}
    \caption{Comparison of NR and 1PAT1 (Sec. \ref{subsec:1PAT1_model}) waveforms for an example simulation. The real part of the $(2,2)$ and $(2,1)$ modes are shown in the top and bottom panels. The binary parameter and SXS identifier of the NR waveform are shown at the top of the plot. The waveforms are aligned in time and phase at the reference time of the NR simulation, which has been shifted to $t=0$. The inset shows a close-up of a region near the merger.}
    \label{fig:q_20_1PAT1_NR}
\end{figure*}

By combining equations (\ref{eq:orb_freq_evol_eq_2}), (\ref{eq:a_of_x}) and (\ref{eq:Omega_of_t}) and then integrating given the initial conditions in Eq. (\ref{eq:initial_conditions}), we calculate the 1PA GW phase and orbital frequency; after feeding this result into equations (\ref{eq:A_1_amp})-(\ref{eq:1PAT1_hlm_resummed}), we are then able to obtain the inspiral waveforms. For the waveform amplitudes, we generate each mode up to $\ell=5$. Although data for both first- and second-order GW fluxes exist up to an orbital radius of $r_0 = 30\,M$, we have chosen the initial orbital radius to be $r_0(t_0)=25\,M$, where $t_0=0$, for the following reason: using Eq. (\ref{eq:Omega_of_t}) we see that this choice of initial orbital radius corresponds to an orbital frequency of $8 \times 10^{-3}$ $M^{-1}$, a value smaller than all the initial orbital frequencies listed in Table \ref{table:sim_list}. Therefore, in this work all of the inspiral waveforms, and hence all of the hybrid waveforms, have an initial orbital frequency of $M \Omega_0 = 8 \times 10^{-3}$.

The numerical integration of Eq. (\ref{eq:orb_freq_evol_eq_2}) is terminated before the orbital radius reaches the innermost stable circular orbit (ISCO), here defined as $r_0 = 6\,M$ \cite{Wald:1984rg, 2021PhRvD.103f4048M}. This is because the ISCO marks the point at which post-adiabatic expansion breaks down—when the binary transitions from inspiral to plunge and merger \cite{2024ScPP...17...56K}. To ensure that each 1PAT1 waveform is cut off at an appropriate time, we terminate the integration of Eq. (\ref{eq:orb_freq_evol_eq_2}) upon reaching  the 1PA breakdown frequency given by Eq. (30) in Ref. \cite{2022PhRvD.106h4061A},
\begin{equation}
    \hat{\Omega}_{\text{break}} \approx \hat{\Omega}_{\text{ISCO}} - 0.026\,\nu^{1/4},
    \label{eq:breakdown_freq}
\end{equation}
where $\hat{\Omega}_{\text{ISCO}} = 6^{-3/2}$ is the Schwarzschild geodesic frequency of the ISCO. In this paper, the final orbital radius $r_f$, defined as $r_f(\nu) = M \hat{\Omega}_{\text{break}}^{-2/3}(\nu)$, ranges from $6.83\,M$ ($q=20)$ to $7.4\,M$ ($q=1$). We note that there is no data available for second-order GW flux below $r_0 = 6.5\,M$. Accordingly, the 1PAT1 waveform model is limited to the inspiral regime because it has not yet been implemented in the transition-to-plunge regime, though recent papers have demonstrated frameworks to build waveform models that extend beyond ISCO into the transition-to-plunge \cite{2024ScPP...17...56K} and merger-ringdown \cite{Kuchler:2025hwx} regimes, pointing the way for the construction of an IMR model that quickly and accurately produces waveforms for asymmetric binaries.

The 1PAT1 waveform model uses two parameters to generate inspiral waveforms: the total mass $M$ and the symmetric mass ratio $\nu$. 
Since each NR waveform is truncated at reference time (see Sec. \ref{subsec:NR}), the Christodoulou masses $m_1$ and $m_2$ measured at reference time are used to define its mass parameters; thus, for consistency, and to achieve a meaningful comparison, we use these same masses to define $M$ and $\nu$ in the 1PAT1 model. 
Fig. \ref{fig:q_20_1PAT1_NR} shows an example of a $q=20$ NR waveform with the corresponding 1PAT1 inspiral waveform. The inspiral waveform has been shifted in time so that its phase matches the NR phase at the reference time of the NR waveform. We find that there is good agreement between the waveforms in the early inspiral, and that the 1PAT1 inspiral begins to dephase close to the ISCO.

\section{Hybridization}
\label{sec:hybrid}

In this section we discuss our setup of the 1PAT1 model and outline our procedure to hybridize NR and second-order GSF
 waveforms describing non-spinning, quasi-circular BBHs. With the exception of our choice of matching window and waveform modes, we follow the same hybridization procedure as Sec. V of Ref. \cite{2019PhRvD..99f4045V}.

\subsection{Choice of matching window}

Before stitching together two waveforms to build a hybrid, we first fix the inertial frame of the waveforms by optimizing over frame and time shifts within a time interval called the matching window. Within this window, the inspiral and NR waveforms should agree with each other to an acceptable degree; deviations between the inspiral and NR waveforms, particularly at late times, may introduce significant biases in the resulting hybrid waveform. Ref. \cite{2023PhRvL.130x1402W} demonstrated that NR waveforms and 1PAT1 inspirals agree well with each other in the late inspiral for all mass ratios, including the equal-mass case. For this reason, we choose a matching window close to merger: unless stated otherwise, we use a three-orbit long matching window whose starting point is located ten orbits before the peak of the $L^2$ norm of the strain modes,
\begin{equation}
    A_{tot}(t) = \sqrt{\sum_{\ell,m} |h_{\ell m}(t)|^2}.
    \label{eq:peak_L2}
\end{equation}
To determine the duration of the matching window in time, we use the phase of the $(2,2)$ mode of the NR waveform.

\begin{figure}
    \centering 
    \makebox[\linewidth][c]{ \includegraphics[width=1.05\linewidth]{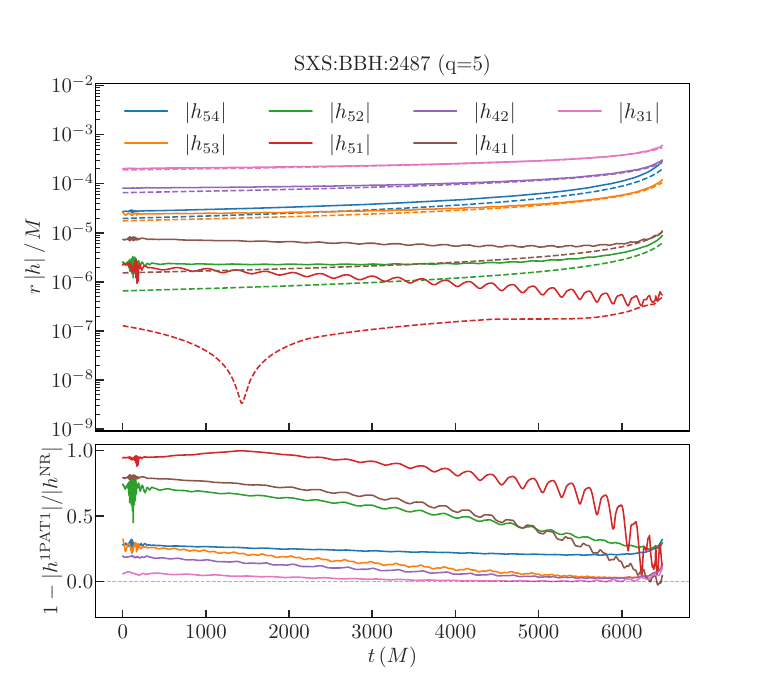}}
    \caption{\textit{Top}: Waveform amplitudes for 1PAT1 (dashed) and NR (solid) for a set of spherical harmonic modes. The binary parameter, $q$, and SXS identifier of the NR waveform are shown at the top of the plot. The waveforms are aligned in time and phase at the reference time of the NR simulation, which has been shifted to $t=0$. We show the modes that have been excluded from the hybridization procedure. \textit{Bottom}: Relative error in the waveform amplitudes for the excluded modes.}
    \label{fig:excluded-modes}
\end{figure}

\begin{figure*}
    \centering
    \makebox[\textwidth][c]{\includegraphics[width=1.1\linewidth]{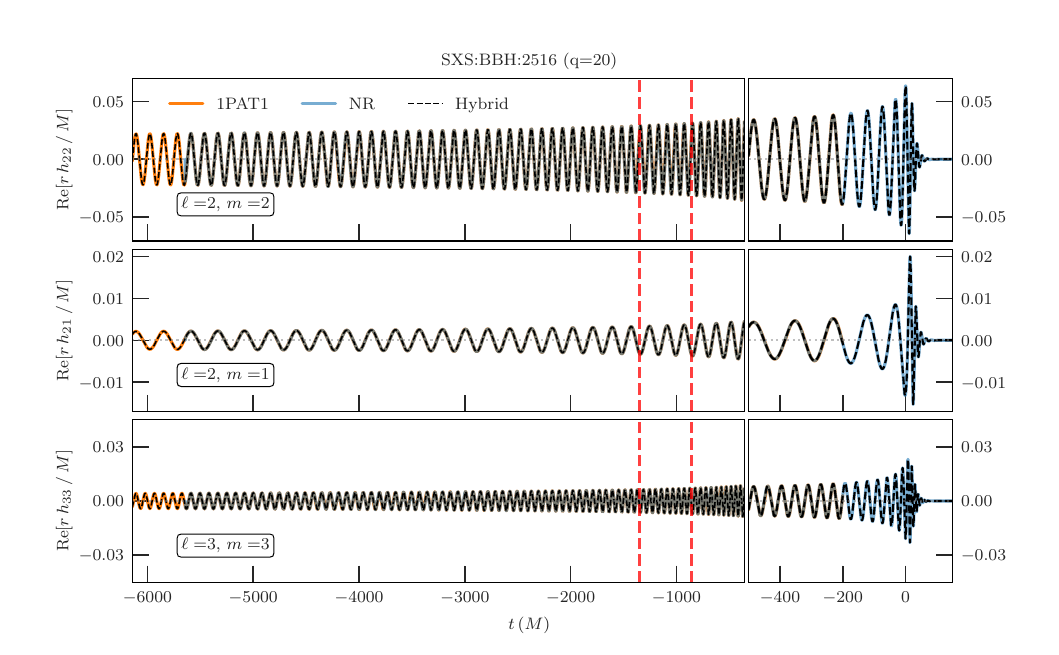}}
    \caption{An example hybrid waveform for a non-spinning $q=20$ simulation. We show the $(2,2)$, $(2,1)$, and $(3,3)$ modes in the top, middle, and bottom rows of the plot. The SXS identifier of the NR waveform is shown at the top of the plot. The vertical dashed red lines represent the matching window. Portions of the waveforms that appear gray are where the 1PAT1 model and NR become indistinguishable. We note that the plot shows the 1PAT1 and NR waveforms after having been transformed into the hybridization frame, and that the time has been shifted such that the peak of the total amplitude occurs at $t=0$.}
    \label{fig:hybrid-modes-q-20}
\end{figure*}

\begin{figure*}
    \centering
    \makebox[\textwidth][c]{\includegraphics[width=1.1\linewidth]{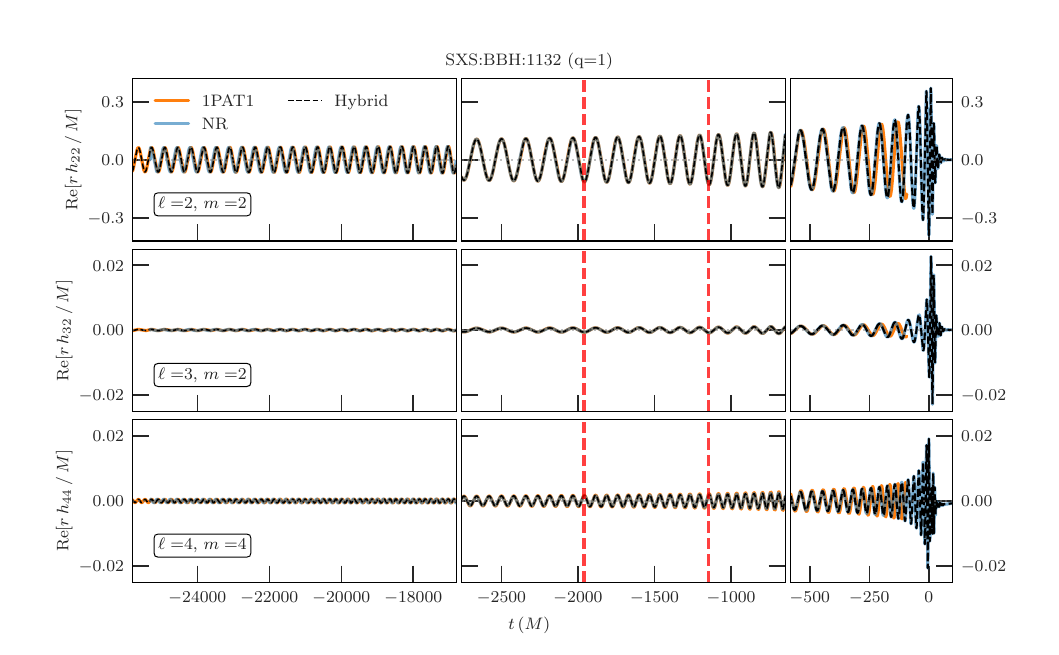}}
    \caption{An example hybrid waveform for a non-spinning $q=1$ simulation. We show the $(2,2)$, $(3,2)$, and $(4,4)$ modes in the top, middle, and bottom rows of the plot. The SXS identifier of the NR waveform is shown at the top of the plot. The vertical dashed red lines represent the matching window. Portions of the waveforms that appear gray are where the 1PAT1 model and NR become indistinguishable. We note that the plot shows the 1PAT1 and NR waveforms after having been transformed into the hybridization frame, and that the time has been shifted such that the peak of the total amplitude occurs at $t=0$. Moreover, because the waveform is long, the plot for each mode has been divided into three sections: early inspiral, matching region, and merger.}
    \label{fig:hybrid-modes-q-1}
\end{figure*}

\subsection{Modes}
\label{subsec:modes}

For the hybrid waveforms, we include the $(2,2)$, $(2,1)$, $(3,3)$, $(3,2)$, $(4,4)$, $(4,3)$, and $(5,5)$ modes. Although the 1PAT1 model can generate inspiral waveforms for modes up to $\ell=5$, we find the amplitudes of the $(3,1)$, $(4,2)$, $(4,1)$, $(5,4)$, $(5,3)$, $(5,2)$, and $(5,1)$ modes for 1PAT1 do not agree, in general, with those of NR, as demonstrated in Fig. \ref{fig:excluded-modes}. In particular, we find that the 1PAT1 amplitudes for the $(5,4)$ and $(5,3)$ modes are smaller than those of NR; this discrepancy becomes more significant for smaller $q$. We also find that, of all the modes, the $(5,1)$ mode has the largest relative error in the waveform amplitude, with the $(4,1)$ and $(5,2)$ modes exhibiting large relative errors as well. Furthermore, while we find that the deviations of 1PAT1 amplitudes from that of NR for the $(3,1)$ and $(4,2)$ modes become less significant with increasing $q$, it is not enough to warrant their inclusion in the aforementioned list of available modes, for they will bias the hybrids at more comparable mass ratios.

After hybridization, we apply a time shift to the hybrid waveforms such that the peak of the $L^2$ norm of the strain modes [see Eq. (\ref{eq:peak_L2})] is aligned at $t=0$. Lastly, we use cubic splines to interpolate the real and imaginary parts of the hybrid waveform modes with a time step of $\Delta t = 0.1\,M$ to ensure a densely-sampled time grid \cite{Virtanen:2019joe}. We present in Figures \ref{fig:hybrid-modes-q-20} and \ref{fig:hybrid-modes-q-1} examples of the final hybrid, NR and 1PAT1 waveforms for mass ratios $q=20$ and $q=1$.

\section{Error Analysis}
\label{sec:results}
In this section we examine the quality of our hybrid waveforms by assessing the error in the hybridization and by performing waveform comparisons with surrogate models.

\subsection{Hybridization errors}
\label{subsec:hybrid-errors}
The creation of hybrid waveforms will introduce errors that need to be assessed to determine their quality. These errors are incurred, for example, by the choice of  hybridization method and inspiral waveform. We employ two methods to estimate these errors based on those used in Ref. \cite{2023PhRvD.108f4027Y}. First, we compare the hybrid waveform to the NR waveform in the matching window; this method allows us to ascertain the errors introduced from the difference between the NR and 2GSF waveforms and from using the smooth transition function in Eq. (28) of Ref.~\cite{2019PhRvD..99f4045V} to stitch the waveforms together within the window. Second, we compare the hybrid to the NR waveform over a time segment in the inspiral portion that extends to the end of the matching window; this check characterizes the error introduced from using 2GSF waveforms for the early inspiral parts of our hybrids. We also include an estimate of the resolution error between the two highest resolution NR waveforms in the matching window. Of the NR simulations listed in Table \ref{table:sim_list}, 46 of them are long enough to be used to estimate the inspiral error. 

To compare two waveforms labeled $h_a$ and $h_b$, where $a$ and $b$ are labels and not Einstein indexes, we use the following cost function $\mathcal{E}$, defined as the normalized $L^2$ difference between the waveforms in  a time segment,
\begin{equation}
    \mathcal{E}[h_a,h_b]=\frac{1}{2} \frac{\sum_{\ell,m} \int_{t_1}^{t_2} |h_{a,\ell m}(t) - h_{b,\ell m}(t)|^2\,dt}{\sum_{\ell,m} \int_{t_1}^{t_2} |h_{a,\ell m}(t)|^2\,dt},
    \label{eq:L2_norm_error}
\end{equation}
 where $t_1$ and $t_2$ denote the start and end of the time segment. This cost function has been shown to reduce to the weighted average of the time-domain mismatch over the sky (see Appendix C of \cite{2017PhRvD..95j4023B}).

\begin{figure}
    \centering
    \includegraphics[width=\linewidth]{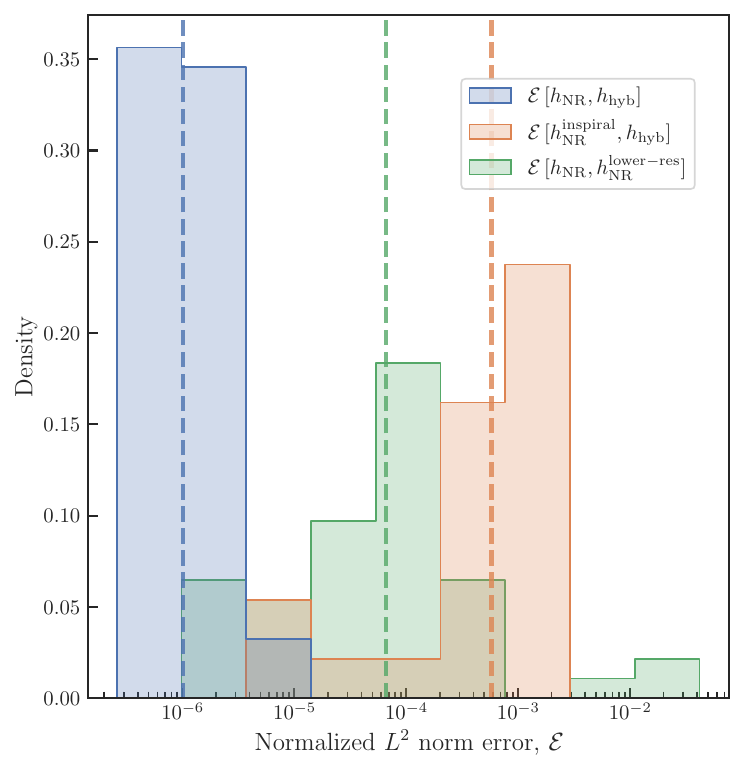}
    \caption{Estimates of the hybridization errors. The measure $\mathcal{E}[h_{\text{NR}}, h_{\text{hyb}}]$, using Eq. (\ref{eq:L2_norm_error}), computes the error between the NR and hybrid waveforms in the matching window. The measure $\mathcal{E}[h_{\text{NR}}^{\text{inspiral}}, h_{\text{hyb}}]$ computes the error between the NR and hybrid waveforms in a time segment with an end time of $t_2$, the end of the matching window, and an initial time of $t_1=t_2-3000\,M$. In addition, we include $\mathcal{E}[h_{\text{NR}}, h_{\text{NR}}^{\text{lower-res}}]$, the resolution error comparing the highest and second-highest resolution NR waveforms within the matching window. While $\mathcal{E}[h_{\text{NR}}, h_{\text{hyb}}]$ was computed for each of the 68 waveforms, $\mathcal{E}[h_{\text{NR}}^{\text{inspiral}}, h_{\text{hyb}}]$ and $\mathcal{E}[h_{\text{NR}}, h_{\text{NR}}^{\text{lower-res}}]$ were computed for the 46 waveforms for which longer and lower-resolution simulations were available; we note that there was a negligble difference in the blue histogram if the additional 22 waveforms were omitted. The histograms are normalized such that the area under each curve when integrated over $\text{log}_{10}(\mathcal{E})$ is 1. The dashed vertical lines represent the median values.}
    \label{fig:hybrid-errors-histogram}
\end{figure}

For the first comparison, we compute the error between a series of hybrid and NR waveforms of a given binary parameter $q$, within the matching window using Eq. (\ref{eq:L2_norm_error}), where $t_1$ and $t_2$ are the start and end time of the window, respectively. For the second comparison, we compute the error between the same set of hybrid and NR waveforms during the insprial, where $t_2$ is the end of the matching window and $t_1 = t_2 - 3000\,M$. The results of these comparisons are shown in Fig. \ref{fig:hybrid-errors-histogram}. We see in Fig. \ref{fig:hybrid-errors-histogram} that the error due to hybridization within the matching window tends to be lower than the NR resolution error, whereas the error during inspiral tends to be higher overall, indicating that a limiting source of error in the hybrid waveforms comes from the 1PAT1 inspiral model. We expect, however, that the inspiral error will decrease for increasing mass ratio as the accuracy of the model improves; in contrast, we expect NR resolution error to increase for increasing mass ratio because of the difficulty in simulating BBH systems in this region of parameter space.

\subsection{Comparison against surrogate models}
In this section we investigate the impact of subdominant modes on the accuracy of the hybrid waveforms. We evaluate the accuracy of the hybrid waveforms by comparing against two surrogate models, \texttt{NRHybSur2dq15} \cite{2022PhRvD.106d4001Y} (valid for $q \leq 15$) and \texttt{BHPTNRSur1dq1e4} \cite{2022PhRvD.106j4025I} (valid for $2.5 \leq q \leq 10^4$). To estimate the difference between two waveforms $h_a$ and $h_b$, we compute the frequency-domain mismatch $\mathcal{M}$ via
\begin{equation}
    \mathcal{M} = 1 - \frac{\langle h_a, h_b \rangle}{\sqrt{\langle h_a, h_a \rangle \langle h_b, h_b \rangle}},
\end{equation}
where $\langle \cdot , \cdot \rangle$ is the frequency domain noise-weighted inner product
\begin{equation}
    \langle h_a, h_b \rangle = 4 \text{Re} \left[ \int_{f_{\text{min}}}^{f_{\text{max}}} \frac{\tilde{h}_a (f) \tilde{h}_b^* (f)}{S_n(f)}\,df \right].
\end{equation}
Here $\tilde{h} (f)$ is the Fourier transform of the complex waveform $h (t)$; * indicates the complex conjugate; Re indicates the real part; and $S_n(f)$ is the one-sided power spectral density of noise in a GW detector. Before computing the mismatches, we first taper both ends of the time-domain waveform using a Planck window \cite{2010CQGra..27h4020M}. The tapering at the start of the waveform is done over the first 1.5 GW cycles of the (2,2) mode, and the tapering at the end is done over the last $20\,M$. We then zero-pad to the nearest power of two.

\begin{figure*}
     \centering
     \begin{subfigure}[]{0.495\textwidth}
         \centering
         \includegraphics[width=\textwidth]{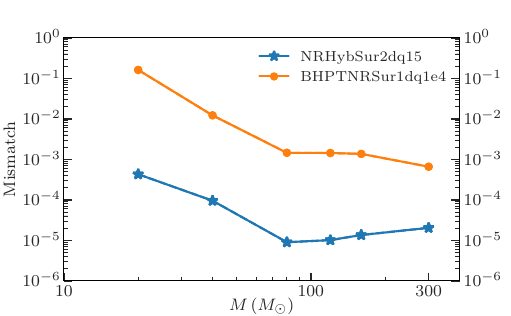}
         \caption{$(2,2)$}
         \label{fig:mismatch_aLIGO_q15_lm_22}
     \end{subfigure}
     %\hfill
     \begin{subfigure}[]{0.495\textwidth}
         \centering
         \includegraphics[width=\textwidth]{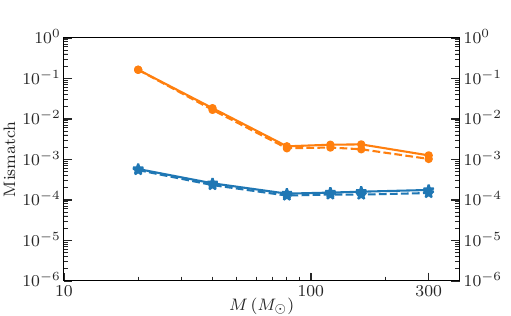}
         \caption{$(2,2), (3,3)$}
         \label{fig:mismatch_aLIGO_q15_lm_22_33}
     \end{subfigure}
     %\hfill
     \begin{subfigure}[]{0.495\textwidth}
         \centering
         \includegraphics[width=\textwidth]{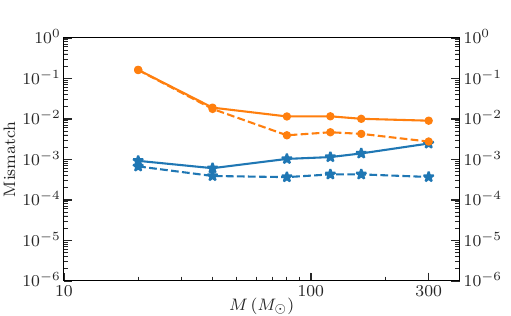}
         \caption{$(2,2), (3,3), (2,1)$}
         \label{fig:mismatch_aLIGO_q15_lm_22_33_21_44}
     \end{subfigure}
     \begin{subfigure}[]{0.495\textwidth}
         \centering
         \includegraphics[width=\textwidth]{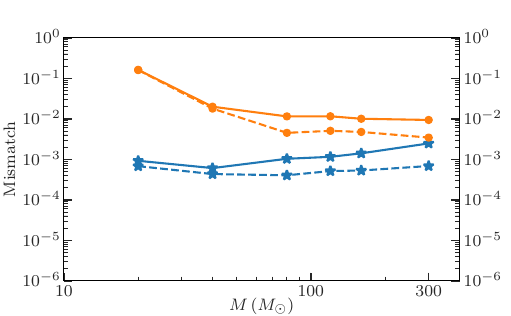}
         \caption{$(2,2), (3,3), (2,1), (4,4), (5,5)$}
         \label{fig:mismatch_aLIGO_q15_lm_22_33_21_44_55}
     \end{subfigure}
        \caption{Frequency-domain mismatches for the $q=15$ hybrid waveform when compared to the \texttt{NRHybSur2dq15} and \texttt{BHPTNRSur1dq1e4} surrogate models. Each panel shows mismatches computed using a subset of shared modes across the hybrid and surrogate waveforms: (a) (2,2); (b) (2,2), (3,3); (c) (2,2), (3,3), (2,1); (d) (2,2), (3,3), (2,1), (4,4), (5,5). Mismatches were computed using the Advanced-LIGO noise curve as a function of total mass over several points in the sky. The solid (dashed) lines represent the 95th percentile (median) mismatch values. We note that in panel (a) the 95th percentile and median mismatch values are indistinguishable by eye.}
        \label{fig:model-comparison-mismatches-q15}
\end{figure*}

\begin{figure*}
     \centering
     \begin{subfigure}[]{0.495\textwidth}
         \centering
         \includegraphics[width=\textwidth]{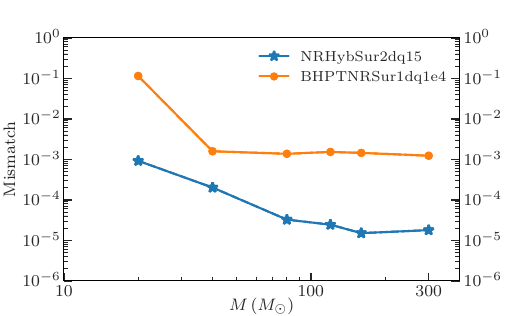}
         \caption{$(2,2)$}
         %\label{fig:y equals x}
     \end{subfigure}
     %\hfill
     \begin{subfigure}[]{0.495\textwidth}
         \centering
         \includegraphics[width=\textwidth]{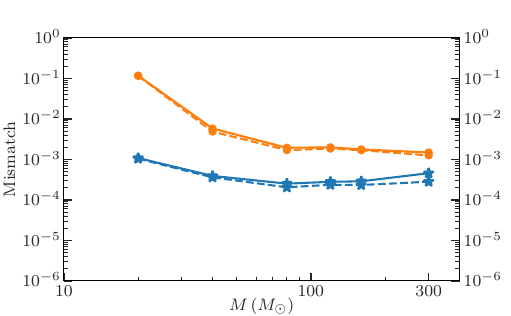}
         \caption{$(2,2), (3,3)$}
         %\label{fig:three sin x}
     \end{subfigure}
     %\hfill
     \begin{subfigure}[]{0.495\textwidth}
         \centering
         \includegraphics[width=\textwidth]{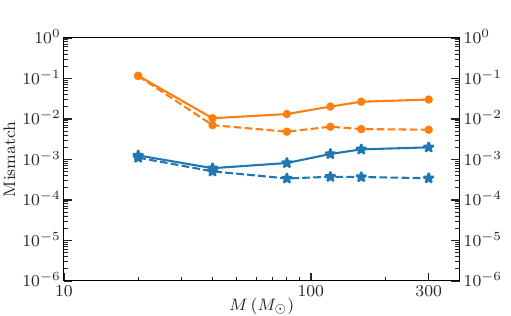}
         \caption{$(2,2), (3,3), (2,1)$}
         %\label{fig:five over x}
     \end{subfigure}
     \begin{subfigure}[]{0.495\textwidth}
         \centering
         \includegraphics[width=\textwidth]{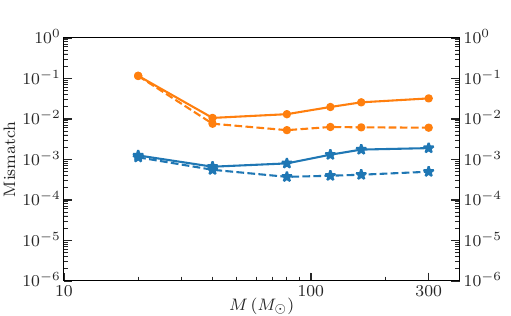}
         \caption{$(2,2), (3,3), (2,1), (4,4), (5,5)$}
         %\label{fig:y equals x}
     \end{subfigure}
        \caption{Same as Fig. \ref{fig:model-comparison-mismatches-q15}, but now showing mismatches for the $q=10$ waveform.}
        \label{fig:model-comparison-mismatches-q10}
\end{figure*}

\begin{figure*}
     \centering
     \begin{subfigure}[]{0.495\textwidth}
         \centering
         \includegraphics[width=\textwidth]{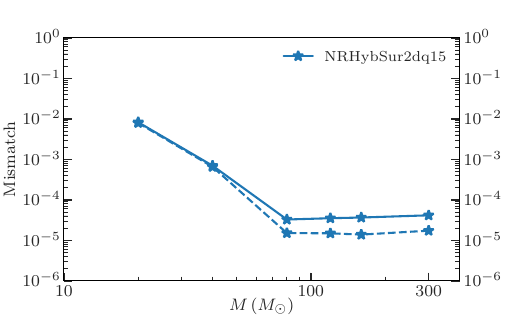}
         \caption{$(2,2)$}
         %\label{fig:three sin x}
     \end{subfigure}
     \begin{subfigure}[]{0.495\textwidth}
         \centering
         \includegraphics[width=\textwidth]{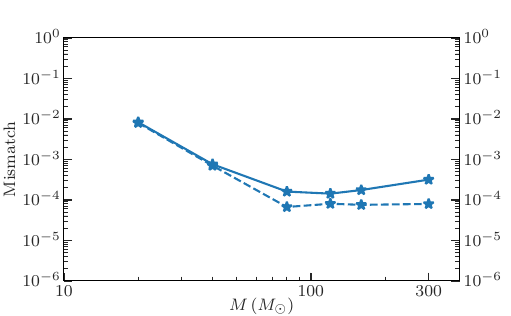}
         \caption{$(2,2), (3,3), (2,1), (4,4), (5,5)$}
         %\label{fig:y equals x}
     \end{subfigure}
        \caption{Same as Fig. \ref{fig:model-comparison-mismatches-q15}, but now showing mismatches for $q=1$ waveforms when compared with the \texttt{NRHybSur2dq15} surrogate model. Because odd-\textit{m} modes vanish for equal mass ratios, we present mismatches for only two sets of modes: one with only the $(2,2)$ mode, and one including all the modes shared between \texttt{NRHybSur2dq15} and the hybrids.}
        \label{fig:model-comparison-mismatches-q1}
\end{figure*}

We optimize the mismatches over shifts in time, polarization angle, and initial orbital phase following the procedure described in Appendix D of Ref. \cite{2017PhRvD..95j4023B}. To treat both plus and cross polarizations on an equal footing, we assume a network of two detectors in which one measures only the plus and the other only the cross polarization. For each pair of waveforms, we compute the mismatches at 37 points uniformly distributed on the sky in the source frame. Lastly, we use the Advanced-LIGO design sensitivity Zero-Detuned-HighP noise curve \cite{aLIGO_noise_curve} with $f_{\text{min}}=20\,\text{Hz}$ and $f_{\text{max}}=2000 \,\text{Hz}$.

By restricting the modes used in computing mismatches to those shared by the model and hybrid waveforms, we can study the error that comes from including subdominant modes in the hybrid waveforms. For this study, we compute mismatches for subsets of the available modes, each of which is drawn from the intersection of available modes for all waveform models: ${(2,2), \, (2,1), \, (3,3), \, (4,4),\text{ and }(5,5)}$. To span a significant range of mass ratios in our set of hybrid waveforms, we compute mismatches for $q=15$, $q=10$, and $q=1$ waveforms. We first compute mismatches using only the dominant $(2,2)$ mode. We then compute mismatches with a new set of modes formed from the union of the previous set and the next leading subdominant mode. This continues iteratively until the full set of shared modes is reached.

Figure \ref{fig:model-comparison-mismatches-q15} shows mismatches computed using the Advanced-LIGO noise curve for \texttt{NRHybSur2dq15} and \texttt{BHPTNRSur1dq1e4} against the $q=15$ hybrid waveform. These mismatches are dependent on total mass, so we show the mismatch for total masses $M$ between $20\,M_{\odot}$ and $300\,M_{\odot}$. At each total mass, we show the median and $95$th percentile mismatches. For each subset of modes, 95th percentile mismatches for \texttt{NRHybSur2dq15} are always below $\sim 3 \times 10^{-3}$ over the entire mass range. When we consider only the $(2,2)$ mode, we find that the mismatches for the \texttt{BHPTNRSur1dq1e4} model are larger than that of \texttt{NRHybSur2dq15} by two orders of magnitude, as shown in Fig. \ref{fig:model-comparison-mismatches-q15}(a); but with the addition of the subdominant modes, the gap between mismatches for both models shrinks to within an order of magnitude for high masses. 

Figure \ref{fig:model-comparison-mismatches-q10}, similarly, shows mismatches against the $q=10$ hybrid waveform. This time 95th percentile mismatches for \texttt{NRHybSur2dq15} are always below $\sim 2 \times 10^{-3}$ over the entire mass range. We also see that \texttt{BHPTNRSur1dq1e4} has mismatches that are at least an order of magnitude larger than that of \texttt{NRHybSur2dq15} for each subset of modes. Furthermore, we note that the addition of the first-leading subdominant mode $(3,3)$ in Fig. \ref{fig:model-comparison-mismatches-q10}(b) leads to the mismatches for \texttt{NRHybSur2dq15} becoming generally uniform across all masses; that the addition of the $(2,1)$ mode leads to an increase in error at high masses, as shown in Fig. \ref{fig:model-comparison-mismatches-q10}(c); and that the remaining subdominant modes exhibit no change in mismatches with their inclusion.

Lastly, Fig. \ref{fig:model-comparison-mismatches-q1} shows mismatches for \texttt{NRHybSur2dq15} against 17 $q=1$ hybrid waveforms. The \texttt{BHPTNRSur1dq1e4} model was excluded from the comparison because it does not generate waveforms below $q=2.5$. In Fig. \ref{fig:model-comparison-mismatches-q1}(a), we see that 95th percentile mismatches for \texttt{NRHybSur2dq15} remain generally constant at $\sim 4 \times 10^{-5}$ for high masses, but then increase for $M < 80\,M_{\odot}$, reaching $\sim 10^{-2}$ at 20 $M_{\odot}$. This is likely because the early inspiral of the hybrid waveforms is constructed from 1PAT1 inspirals, which exhibit more dephasing as $q \to 1$. We note that the inclusion of the $(4,4)$ mode in Fig. \ref{fig:model-comparison-mismatches-q1}(b) results in an increase in errors for $M \geq 80\,M_{\odot}$, whereas errors for low masses exhibit no change.

\begin{figure*}
    \centering
    \includegraphics[width=1.0\linewidth]{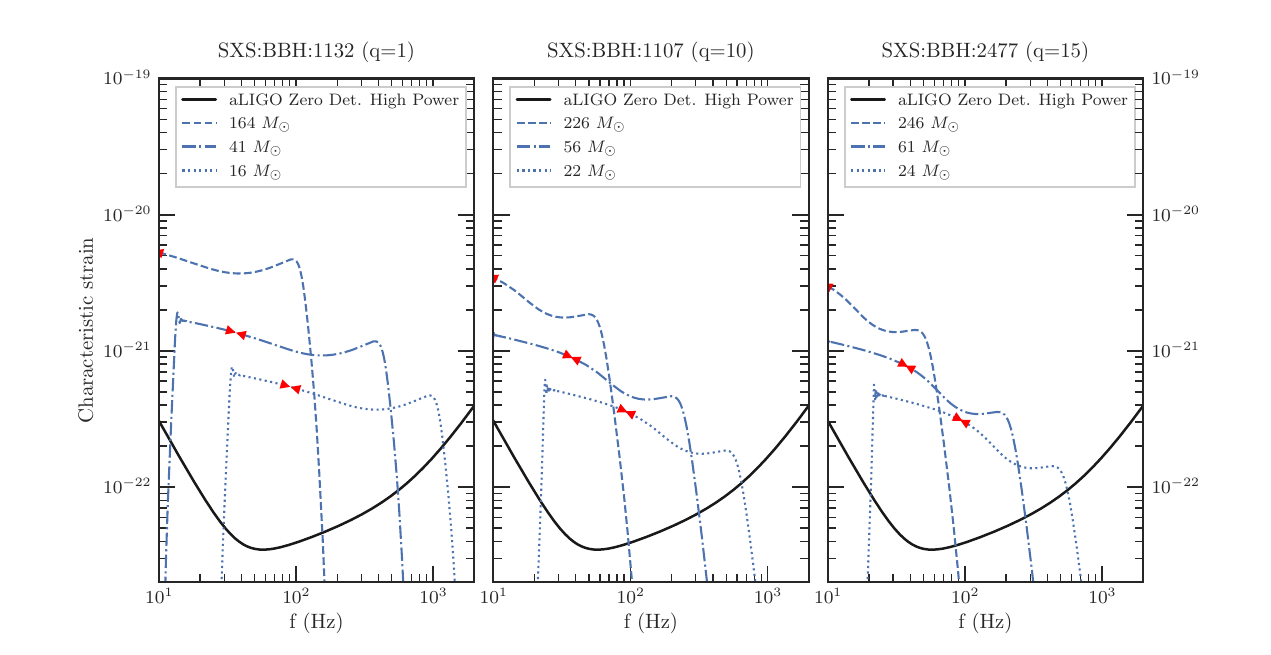}
    \caption{Characteristic strain of hybrid waveforms for mass ratios $q=1$, $q=10$, and $q=15$ scaled to various total masses and shown against the Zero-Detuned-HighP noise curve. The red triangles represent the starting and ending frequencies of the matching window. The total mass of each binary is scaled such that the matching window ends at 10 Hz, 40 Hz, and 100 Hz. The amplitude of a given waveform is scaled such that it represents a binary at a distance of 1 Gpc from the detector.}
    \label{fig:char-strain-vs-PSD}
\end{figure*}

To better understand how the 1PAT1 inspiral model affects mismatches between hybrid waveforms and the surrogate waveforms, we supplement the mismatches in Figs. \ref{fig:model-comparison-mismatches-q15}, \ref{fig:model-comparison-mismatches-q10}, and \ref{fig:model-comparison-mismatches-q1} with Fig. \ref{fig:char-strain-vs-PSD}, which shows the characteristic strains for hybrid waveforms with $q=1$, $q=10$, and $q=15$ as well as the Advanced-LIGO noise curve. For instance, at high masses only the NR portion of the hybrid waveforms is in the LIGO frequency band, resulting in relatively lower mismatches when compared to those at low masses, where the 1PAT1 inspiral portion before the matching window starts to enter the LIGO band. For low masses, then, the mismatches are an indicator of how well the surrogate models reproduce the 2GSF-NR hybrid waveforms. We note, though, that the inspiral portion of these hybrids can be further lengthened to span the full frequency band for low masses by increasing the initial orbital radius $r_0$, so long as there are GW flux and redshift data available at these radii.

\section{Search for optimal window location}
\label{sec:search_opt_window}
The need to mitigate the error introduced by hybridization  places restrictions on the choice of the matching window. Generally, the matching window should be chosen as early in the waveform as possible to avoid matching near merger where the inspiral model breaks down, and it should be made as long as possible to obtain a better fit of the data (short windows run the risk of overfitting) \cite{2024PhRvD.110j4076S}.

Unfortunately, we run into a problem when considering high-mass-ratio hybrids: the large computational cost of running NR simulations for either large separations or high mass ratios implies that high-mass-ratio simulations must be short if they are to be completed within a reasonable time frame. This calls into question the viability of producing such hybrids, for the NR waveforms would not be long enough to provide sufficient data for hybridization. However, this uncertainty arises because of the assumption that the inspiral model becomes inaccurate well before merger occurs. As demonstrated in Ref. \cite{2023PhRvL.130x1402W}, the 1PAT1 waveform model agrees with NR up to, but not including, the transition to plunge, and this agreement is expected to improve with increasing mass ratio. Consequently, if we can place the matching window as close to merger as possible without introducing significant error in a 2GSF-NR hybrid waveform, we can cut down on computational costs by launching high mass ratio NR simulations at smaller separations and using hybridization with 1PAT1 inspirals to lengthen the NR waveforms.

We investigate this possibility by determining whether or not there is a correlation between mass ratio and window placement for hybrid waveforms made with 1PAT1 inspirals. We proceed as follows. First, we place the window such that its right end is aligned with $t=0$. Second, for each choice of window, we use Eq. (\ref{eq:L2_norm_error}) to quantify the inspiral error—the measure $\mathcal{E}[h_{\text{NR}}^{\text{inspiral}}, h_{\text{hyb}}]$ used in Sec. \ref{subsec:hybrid-errors}—between the hybrid and NR waveforms. Third, if the inspiral error is larger than a given tolerance level, we shift the window leftward towards the early inspiral by one GW cycle. We iterate the last two steps until the inspiral error has dropped below the tolerance level, upon which we designate the current location of the window as the optimal choice. If the window reaches the start of the NR waveform before the criterion is met, the data point is excluded; similarly, hybrid waveforms for which the end point of the 1PAT1 inspiral is within the matching window are also excluded (otherwise we would get unphysical glitches in residuals between the hybrid and NR waveforms near the window's ending point).  For this study we fix the length of the window to 6 GW cycles, and we set the tolerance level to $10^{-3}$.

\begin{figure}
    \centering
    \includegraphics[width=\linewidth]{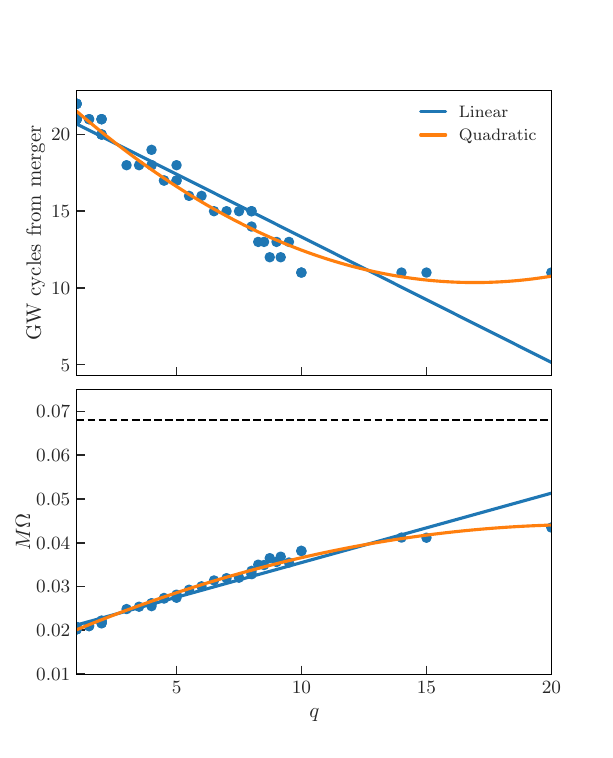}
    \caption{\textit{Top}: Optimal locations of the matching window for hybrid waveforms with mass ratios in the range $1 \leq q \leq 20$. We vary the starting time (shown in number of GW cycles from merger) of the matching window, while keeping the window length fixed at 6 GW cycles, until the inspiral error falls under $10^{-3}$. There are a total of 51 hybrids that meet this condition. The blue (orange) curve is a linear (quadratic) fit to the plotted points. \textit{Bottom}: Same as top panel, but now showing the optimal starting orbital frequency of the matching window versus mass ratio. The horizontal dashed line represents $\hat{\Omega}_{\text{ISCO}} = 6^{-3/2}$, the Schwarzschild geodesic frequency of the ISCO.}
    \label{fig:q_vs_cycles_and_orb_freq}
\end{figure}

\begin{figure}
    \centering
    \includegraphics[width=1.0\linewidth]{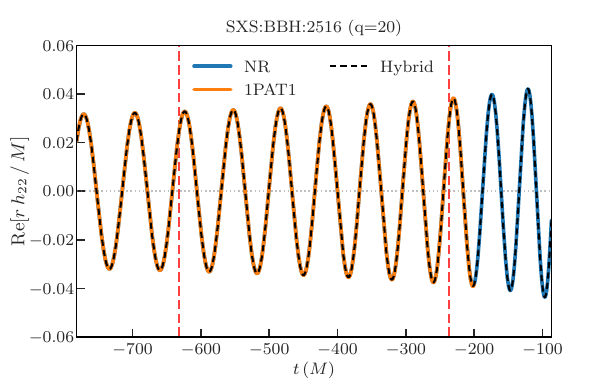}
    \caption{The real part of the (2,2) mode of 1PAT1, NR, and hybrid waveforms. The binary parameter and SXS identifier of the NR waveform are shown at the top of the plot. The vertical dashed red lines represent the matching window. We note that the 1PAT1 and NR waveforms shown in this plot have been transformed into the hybridization frame (see Sec. V of Ref. \cite{2019PhRvD..99f4045V}); and that the time has been shifted such that merger occurs at $t=0$. The optimal location of the start of this window is 11 GW cycles from merger, the same as the $q=14$ and $q=15$ hybrid waveforms.}
    \label{fig:optimal-window-q20}
\end{figure}

In Figure \ref{fig:q_vs_cycles_and_orb_freq} we plot the optimal locations of the matching window for various mass ratios. The vertical axis for the top panel denotes the location of the start of the matching window relative to merger ($t=0$) in number of GW cycles. Of the simulations in our dataset, 51 of them meet the condition we imposed on the inspiral error. We observe that the blue points in Fig. \ref{fig:q_vs_cycles_and_orb_freq} reveal a correlation between the placement of the window and the mass ratio of an NR simulation: namely, the window shifts towards merger with increasing mass ratio. To better understand this correlation, we perform linear and quadratic fits to the points, represented by blue and orange curves. We note that, because of the inclusion of the $q=14$, $q=15$, and $q=20$ simulations in our dataset, the quadratic fit deviates significantly from the linear fit, and flattens out gradually towards $q=20$. The window appears to come to a standstill for $q > 10$. By studying the region around the matching window for each of these three data points, we find that the end point of the window is within one GW cycle of the end point of the inspiral waveform, as shown in Fig. \ref{fig:optimal-window-q20}. This suggests that the placement of the matching window for constructing high-mass-ratio hybrids is limited by the 1PA breakdown frequency described in Eq. (\ref{eq:breakdown_freq}).

With the knowledge that we can place the window near merger for large mass ratios—at least according to our criterion for choosing the window location—we now consider the initial orbital frequencies of NR waveforms for which hybridization with 1PAT1 inspirals can take place. To help us determine how large the initial orbital frequency—and conversely how small the initial binary separation—of NR simulations can be for such hybridization, in the bottom panel of Fig. \ref{fig:q_vs_cycles_and_orb_freq} we plot the optimal starting orbital frequency of the matching window against mass ratio. We note that the orange curve represents an upper bound for the maximum initial orbital frequency of an NR simulation with mass ratio $q$; because of initial transients in NR simulations of BBHs, however, the curve overestimates this upper bound.

\section{Conclusions}
\label{sec:conclusions}

In this work we present 2GSF-NR hybrid waveforms, the inspiral of which comes from the 1PAT1 model. These hybrid waveforms are generated from 68 NR waveforms representing non-spinning, quasi-circular BBH systems with mass ratios $q \leq 20$, and include the $(2,2)$, $(2,1)$, $(3,3)$, $(3,2)$, $(4,4)$, $(4,3)$ and $(5,5)$ spin-weighted spherical harmonic modes.

By comparing the hybrid waveforms against the NR waveforms in our dataset, we show that the errors in our hybridization procedure are comparable to, or lower than, NR resolution errors, with the caveat that the resolution errors were computed by comparing the NR waveforms with those from lower-resolution simulations. We also perform a mismatch comparison between hybrid waveforms and surrogate waveforms using different sets of modes for $q=1$, $q=10$, and $q=15$. We find that, for higher $q$, \texttt{NRHybSur2dq15} reproduces the hybrid waveforms with mismatches $\sim 3 \times 10^{-3}$ for total masses in the range $20\, M_{\odot} \leq M \leq 300 \, M_{\odot}$. For $q=1$, however, mismatches increase to $\sim 10^{-2}$ at low masses owing to dephasing in the 1PAT1 model at comparable-mass ratios.

Through a study of the placement of the matching window, we show that the window can be set closer to merger for higher mass ratios without introducing significant error in the late inspiral of hybrid waveforms. We find, however, that the start of the window remains at 11 GW cycles for $q > 10$; the window cannot be placed any closer to merger because it reaches the end point of the 1PAT1 inspiral, which depends on the symmetric mass ratio—and, therefore, mass ratio—of the system. Our results suggest, then, that the duration of high-mass-ratio NR simulations launched for the purposes of hybridization with 1PAT1 inspirals will be constrained by the 1PA breakdown frequency.

In the future, we will study how the placement of the matching window in hybrid injections impacts the recovery of source parameters. This study would facilitate the investigation of the feasibility of producing accurate hybrid waveforms from high-mass-ratio NR simulations launched at short separations.

\acknowledgements
We thank Niels Warburton, Adam Pound, Aaron Zimmerman, and Jonathan Thompson for insightful discussions on this project. HI additionally thanks Niels Warburton for granting him access to an updated version of the 1PAT1 model. We thank Aasim Jan for sharing his mismatch code with us. Mismatch calculations were performed on the CIT cluster of the LIGO Laboratory, which is supported by NSF Grants PHY-0757058 and PHY-0823459. This research was supported by NASA grant No. 80NSSC24K0437 and NSF PHY-2207780. This work makes use of public NR data from the SXS Collaboration \cite{Scheel:2025jct}; the Black Hole Perturbation Toolkit \cite{BHPToolkit}; Python libraries including \texttt{SciPy} \cite{Virtanen:2019joe}, \texttt{NumPy} \cite{Harris:2020xlr}, \texttt{matplotlib} \cite{Hunter:2007}, and \texttt{seaborn} \cite{Waskom2021}; and the \texttt{gwsurrogate} \cite{2014PhRvX...4c1006F, gwsurrogatepypi} and \texttt{gwtools} \cite{gwtoolspypi} packages. Furthermore, this work was done by members of the Weinberg Institute and has an identifier of WI Preprint UT-WI-11-2025.

\appendix

\section{NR Simulations}
\label{appendix:NR}

We use the following dataset from the third SXS public catalog (Table \ref{table:sim_list}) to hybridize NR waveforms with 1PAT1 inspirals \cite{Scheel:2025jct}. These simulations were chosen because (i) their mass ratios span from $q=1$ to $q=20$ and (ii) they are initially slowly spinning.

\begin{table*}
\begin{minipage}{\columnwidth}
    \centering
    \begin{tabular}{p{.26\columnwidth} p{.07\columnwidth} p{.155\columnwidth} p{.155\columnwidth} p{.11\columnwidth}  p{.155\columnwidth} }
      \hline \hline \\ %inserts double horizontal lines
SXS ID  & $q$ & $M \Omega_0$ & $M \omega_{\text{ref}}$ & $N_{\text{orbits}}$ & $e_{\text{ref}}$ \\ [0.5ex] % inserts table
%heading 0001
\hline  % inserts single horizontal line
SXS:BBH:0389 & 1 & 0.015246 & 0.016039 & 18.6 & 8.460e-05 \\ % inserting body of the table
SXS:BBH:2377 & 1 & 0.012226 & 0.012472 & 28.0 & 1.534e-04 \\
SXS:BBH:2378 & 1 & 0.012227 & 0.012699 & 28.0 & 1.079e-04 \\
SXS:BBH:3624 & 1 & 0.012225 & 0.012348 & 28.0 & 1.465e-04 \\
SXS:BBH:2598 & 1 & 0.012623 & 0.012511 & 27.1 & 4.029e-03 \\
SXS:BBH:4434 & 1 & 0.015335 & 0.017482 & 18.4 & 1.289e-04 \\
SXS:BBH:2496 & 1 & 0.014457 & 0.014683 & 20.9 & 1.615e-04 \\
SXS:BBH:1132 & 1 & 0.008518 & 0.008540 & 53.4 & 2.700e-05  \\ 
SXS:BBH:3864 & 1 & 0.015336 & 0.016541 & 18.4 & 7.959e-04 \\
SXS:BBH:1154 & 1 & 0.009934 & 0.010211 & 40.7 & 5.750e-05  \\
SXS:BBH:1155 & 1 & 0.009934 & 0.010148 & 40.7 & 6.900e-06 \\
SXS:BBH:1153 & 1 & 0.009930 & 0.010214 & 40.6 & 7.651e-04  \\
SXS:BBH:3617 & 1 & 0.011928 & 0.012269 & 29.2 & 2.366e-04 \\
SXS:BBH:2375 & 1 & 0.012226 & 0.012470 & 28.0 & 1.011e-04 \\
SXS:BBH:2376 & 1 & 0.012226 & 0.012574 & 28.0 & 1.028e-04  \\
SXS:BBH:2325 & 1 & 0.012226 & 0.012470 & 28.0 & 7.960e-05 \\
SXS:BBH:2326 & 1 & 0.011312 & 0.011702 & 32.2 & 1.668e-04 \\

SXS:BBH:0198 & 1.2 & 0.014483 & 0.015479 & 20.7 & 2.023e-04  \\
SXS:BBH:2331 & 1.5 & 0.012222 & 0.012901 & 29.0 & 5.410e-05  \\
SXS:BBH:3984 & 1.5 & 0.012259 & 0.012239 & 29.2 & 1.863e-04  \\
SXS:BBH:0593 & 1.5 & 0.015451 & 0.016627 & 18.8 & 5.750e-05 \\
%SXS:BBH:0194 & 1.52 & 0.015150 & 0.015570 & 19.6 & 8.020e-04 \\

%SXS:BBH:0238 & 2 & 0.011171 & 0.011260 & 32.0 & 6.400e-05 \\
%SXS:BBH:0554 & 2 & 0.015808 & 0.016596 & 19.3 & 9.750e-05 \\
SXS:BBH:1222 & 2 & 0.012722 & 0.013342 & 28.8 & 6.110e-05 \\
SXS:BBH:2497 & 2 & 0.015299 & 0.015384 & 20.7 & 2.263e-04 \\
SXS:BBH:1166 & 2 & 0.010555 & 0.010647 & 40.4 & 3.799e-04 \\
SXS:BBH:1164 & 2 & 0.010549 & 0.010793	 & 40.3 & 1.297e-03 \\
SXS:BBH:1167 & 2 & 0.010556 & 0.010769 & 40.5 & 3.617e-04 \\
SXS:BBH:1165 & 2 & 0.010549 & 0.010788 & 40.3 & 1.438e-03 \\

SXS:BBH:0201 & 2.32 & 0.015856 & 0.018710 & 20.0 & 1.938e-04 \\
%SXS:BBH:0259 & 2.5 & 0.013350 & 0.013461 & 28.6 & 4.900e-05 \\

SXS:BBH:1178 & 3 & 0.018952 & 0.021248 & 15.7 & 1.395e-04 \\
SXS:BBH:1179 & 3 & 0.018952 & 0.019860 & 15.7 & 3.690e-05 \\
SXS:BBH:1177 & 3 & 0.018939 & 0.021362 & 15.6 & 2.685e-03 \\
SXS:BBH:2498 & 3 & 0.016589 & 0.019273 & 20.4 & 1.642e-04 \\
SXS:BBH:2265 & 3 & 0.008812 & 0.008915 & 65.6 & 7.170e-05 \\
%SXS:BBH:0200 & 3.27 & 0.016949 & 0.017302 & 20.2 & 4.137e-04 \\
SXS:BBH:2483 & 3.5 & 0.014568 & 0.015706 & 27.8 & 2.336e-04 \\ [1ex]

\hline %inserts single line
\hline

    \end{tabular}
\end{minipage}\hfill % maximize the horizontal separation
\begin{minipage}{\columnwidth}
    \centering
    \begin{tabular}{p{.26\columnwidth} p{.07\columnwidth} p{.155\columnwidth} p{.155\columnwidth} p{.11\columnwidth}  p{.155\columnwidth} }
      \hline \hline \\ %inserts double horizontal lines
SXS ID  & $q$ & $M \Omega_0$ & $M \omega_{\text{ref}}$ & $N_{\text{orbits}}$ & $e_{\text{ref}}$ \\ [0.5ex] % inserts table
%heading 0001
\hline  % inserts single horizontal line
SXS:BBH:2485 & 4 & 0.015700 & 0.016831 & 25.7 & 9.793e-04 \\
SXS:BBH:3631 & 4 & 0.020311 & 0.021177 & 15.4 & 2.058e-04 \\
%SXS:BBH:1906 & 4 & 0.017645 & 0.018080 & 20.4 & 1.472e-04 \\
SXS:BBH:2499 & 4 & 0.017880 & 0.018073 & 20.2 & 1.455e-04 \\
SXS:BBH:1220 & 4 & 0.015525 & 0.016285 & 26.3 & 1.099e-04 \\
SXS:BBH:2484 & 4.5 & 0.015700 & 0.016444 & 27.4 & 8.098e-04 \\
%SXS:BBH:0295 & 4.5 & 0.015665 & 0.015765 & 27.8 & 2.670e-05 \\
SXS:BBH:3144 & 4.5 & 0.014213 & 0.015047 & 33.3 & 2.219e-04 \\

SXS:BBH:2374 & 5 & 0.017413 & 0.017955 & 23.7 & 3.260e-05 \\
SXS:BBH:2487 & 5 & 0.015696 & 0.016647 & 29.1 & 2.376e-04 \\
SXS:BBH:3619 & 5 & 0.015795 & 0.015974 & 28.8 & 7.360e-05 \\
%SXS:BBH:0296 & 5.5 & 0.016589 & 0.016677 & 27.9 & 3.300e-05 \\
SXS:BBH:2486 & 5.5 & 0.015694 & 0.016598 & 30.8 & 4.536e-04 \\

SXS:BBH:2164 & 6 & 0.019260 & 0.021078 & 21.4 & 4.762e-04 \\
SXS:BBH:2489 & 6 & 0.015694 & 0.016509 & 32.6 & 5.865e-04 \\
SXS:BBH:3630 & 6 & 0.019315 & 0.020410 & 21.2 & 4.340e-04 \\
SXS:BBH:2488 & 6.5 & 0.015695 & 0.016530 & 34.4 & 7.904e-04 \\
%SXS:BBH:0297 & 6.5 & 0.020679 & 0.020821 & 19.7 & 5.900e-05 \\
SXS:BBH:0192 & 6.58 & 0.019991 & 0.021679 & 21.1 & 4.350e-05 \\

%SXS:BBH:0298 & 7 & 0.021154 & 0.021297 & 19.7 & 4.000e-05 \\
SXS:BBH:2491 & 7 & 0.015692 & 0.016496 & 36.2 & 3.607e-04 \\
%SXS:BBH:1110 & 7 & 0.006793 & 0.006785 & 175.7 & 3.151e-04 \\
SXS:BBH:0188 & 7.19 & 0.019984 & 0.023439 & 22.3 & 1.504e-04 \\
SXS:BBH:2490 & 7.5 & 0.015686 & 0.016467 & 37.9 & 5.635e-04 \\
%SXS:BBH:0299 & 7.5 & 0.021398 & 0.021525 & 20.1 & 5.600e-05 \\
SXS:BBH:0195 & 7.76 & 0.020665 & 0.025001 & 21.9 & 2.061e-04 \\

SXS:BBH:2493 & 8 & 0.015682 & 0.016427 & 39.5 & 7.760e-04 \\
SXS:BBH:2707 & 8 & 0.016313 & 0.017111 & 36.5 & 3.110e-04 \\
SXS:BBH:3622 & 8 & 0.019293 & 0.019713 & 25.8 & 3.758e-04 \\
SXS:BBH:0186 & 8.27 & 0.020442 & 0.020625 & 23.7 & 1.529e-04 \\
%SXS:BBH:0300 & 8.5 & 0.022972 & 0.023109 & 18.7 & 6.000e-05 \\
SXS:BBH:2492 & 8.5 & 0.015684 & 0.016373 & 41.3 & 8.259e-04 \\
SXS:BBH:0199 & 8.73 & 0.021133 & 0.021513 & 22.6 & 5.470e-05 \\

%SXS:BBH:0301 & 9 & 0.023253 & 0.023380 & 18.9 & 5.700e-05 \\
SXS:BBH:2495 & 9 & 0.015687 & 0.016358 & 43.2 & 1.877e-04 \\
SXS:BBH:0189 & 9.17 & 0.020419 & 0.020685 & 25.2 & 1.020e-04 \\
SXS:BBH:1108 & 9.2 & 0.019345 & 0.022808 & 28.7 & 9.160e-05 \\
SXS:BBH:2494 & 9.5 & 0.015683 & 0.015896 & 44.9 & 1.652e-04 \\
%SXS:BBH:0302 & 9.5 & 0.023539 & 0.023656 & 19.1 & 5.400e-05 \\
%SXS:BBH:0196 & 9.66 & 0.021621 & 0.021782 & 23.1 & 2.629e-04 \\

SXS:BBH:0185 & 9.99 & 0.021124 & 0.021418 & 24.9 & 2.772e-04 \\
SXS:BBH:1107 & 10 & 0.019337 & 0.019575 & 30.4 & 1.051e-03 \\
%SXS:BBH:0303 & 10 & 0.023830 & 0.023946 & 19.3 & 5.600e-05 \\ 

SXS:BBH:2480 & 14 & 0.022602 & 0.022921 & 27.7 & 3.225e-04 \\
SXS:BBH:2477 & 15 & 0.023045 & 0.024128 & 27.9 & 3.746e-04 \\
SXS:BBH:2516 & 20 & 0.023211 & 0.024418 & 34.4 & 1.864e-04 \\ [1ex] % [1ex] adds vertical space

\hline %inserts single line
\hline
    \end{tabular}
  \end{minipage}
  \caption{Configurations for the 68 NR simulations from the third SXS catalog used in this hybridization study. Column 2 shows the mass ratio of the simulation at the reference time. Columns 3 and 4 show the initial orbital frequency and the magnitude of the orbital angular frequency at the reference time. Column 5 shows the number of orbits until the formation of a common apparent horizon, and column 6 shows the eccentricity at the reference time.} % title of Table
\label{table:sim_list} % is used to refer this table in the text
\end{table*}

\section{A closer look at the 1PAT1 model}
\label{appendix:1PAT1}

By combining equations (\ref{eq:E_SF}), (\ref{eq:orb_freq_evol_eq_2}), (\ref{eq:a_of_x}) and (\ref{eq:Omega_of_t}) with the expressions for $x$, $\hat{\Omega}$, $F_0(x)$ and $F_1(x)$, we can derive an expanded form of the frequency evolution of the 1PA GW orbital frequency, $\Omega_{\text{1PA}}(t)$, as a function of $r_0(t)$,
\begin{widetext}
    \begin{align}
        \dv{\Omega_{\text{1PA}}}{t} &= \frac{3 \left(1-\frac{3M}{r_0(t)}\right)^{3/2} \sqrt{\frac{M}{r_0(t)}}}{M^2 \left(1-\frac{6M}{r_0(t)}\right)} \mathcal{F}^1_\nu (r_0(t))\nu \nonumber \\
        & \quad + \frac{3\left(1-\frac{3M}{r_0(t)}\right)^{3/2} \sqrt{\frac{M}{r_0(t)}}}{M^2 \left(1-\frac{6M}{r_0(t)}\right)}\left \{ 2 \frac{\left(1-\frac{3M}{r_0(t)}\right)^{3/2}}{\left(1-\frac{6M}{r_0(t)}\right)} \mathcal{F}^1_\nu (r_0(t)) \partial_x E^{1^{st} \text{law}} (x) + \mathcal{F}^2_\nu (r_0(t)) \right \} \nu^2, 
        \label{eq:freq_evol_expanded}
    \end{align}
\end{widetext}
where the derivative of $E^{1^{st} \text{law}}$ is
\begin{align}
    \partial_x E^{1^{st} \text{law}} (x) &= \frac{z_{\text{SF}}'(x)}{6} - \frac{1}{3}x z_{\text{SF}}''(x) - \frac{3}{2}\frac{1}{\sqrt{1-3x}} \nonumber \\ 
    & + \frac{7-24x}{6(1-3x)^{3/2}} + \frac{3}{4}\frac{(7-24x)x}{(1-3x)^{5/2}} - \frac{4x}{(1-3x)^{3/2}}.
    \label{eq:partial_x_E}
\end{align}
Since we had data for the first-law binding energy as a function of $x$, we applied the chain rule in Eq. (\ref{eq:orb_freq_evol_eq_2}) to get a derivative of $\hat{E}_{\text{SF}} \approx E^{1^{st} \text{law}}$ with respect to $x$. We note that the 0PA GW orbital frequency can be determined similarly by discarding the $\nu^2$ term in Eq. (\ref{eq:freq_evol_expanded}).

The GW amplitudes are determined from 
\begin{align}
 h_{l m}^{(1)}(\Omega_{\text{1PA}}(t))&=-\frac{2 Z_{l m}^{1, \infty}\left([M\, \Omega_{\text{1PA}}(t)]^{-2 / 3}\right)}{(m M \Omega_{\text{1PA}}(t))^2} \nonumber \\ 
 &= -\frac{2}{m^2}\left(\frac{r_0(t)}{M}\right)^3\,Z_{\ell m}^{1,\infty}\left(\frac{r_0(t)}{M}\right), \label{eq:h_1_final}\\ \nonumber \\
 h_{l m}^{(2)}(\Omega_{\text{1PA}}(t))&=\frac{2 Z_{l m}^{2, \infty}\left([M \, \Omega_{\text{1PA}}(t)]^{-2 / 3}\right)}{(m M \Omega_{\text{1PA}}(t))^2} \nonumber \\ 
 &= \frac{2}{m^2}\left(\frac{r_0(t)}{M}\right)^3\,Z_{\ell m}^{2,\infty}\left(\frac{r_0(t)}{M}\right),
\label{eq:h_2_final}
\end{align}
where $Z_{\ell m}^{1,\infty}$ and $Z_{\ell m}^{2,\infty}$ are the first- and second-order radially perturbed Weyl scalars evaluated at infinity, which can be calculated using the BHPToolkit \cite{BHPToolkit, PhysRevLett.29.1114}. After inserting equations (\ref{eq:h_1_final}) and (\ref{eq:h_2_final}) into (\ref{eq:A_1_amp}) and (\ref{eq:A_2_amp}), we find that the GW amplitudes at 0PA and 1PA order are
\begin{align}
    A_{\ell m}^1 (t) &= h_{\ell m}^{(1)}(\Omega_{\text{1PA}}(t)), \\ \nonumber \\
    A_{\ell m}^2 (t) &= h_{\ell m}^{(1)}(\Omega_{\text{1PA}}(t)) + h_{\ell m}^{(2)}(\Omega_{\text{1PA}}(t)) \nonumber \\
    & \quad + \frac{2 r_0(t)}{3 M} \left( 1-\frac{\frac{M}{r_0(t)}}{\sqrt{1-\frac{3 M}{r_0(t)}}} \right) \dv{}{r_0}  h_{\ell m}^{(1)}(\Omega_{\text{1PA}}(t)).
\end{align}\\

\bibliographystyle{ieeetr}
\bibliography{references}

\end{document}